\documentclass[preprint,tightenlines,a4paper,11pt,aps,amsmath,amssymb,showkeys,groupedaddress]{revtex4}

\usepackage{dcolumn}
\usepackage{graphicx, epsfig, color}
\usepackage{amsmath,amssymb}
\usepackage{bm}
\usepackage{multirow}
\def\be{\begin{equation}}
\def\ee{\end{equation}}
\def\bea{\begin{eqnarray}}
\def\eea{\end{eqnarray}}
\def\beas{\begin{eqnarray*}}
\def\eeas{\end{eqnarray*}}
\begin{document}

\title{Synchronization of Four Non-identical Local Coupled Phase Oscillators in A Ring: Analytic Determination of The Critical Coupling and Phase Differences}
\author{Hassan F. El-Nashar{\footnote{corresponding author}}}
\email[]{hfelnashar@sci.asu.edu.eg}
\affiliation{Physics Department, Faculty of Science, Ain Shams University, Cairo 11566, Egypt.}

\author{M. S. Mahmoud}
\email[]{mohamedshehata@sci.asu.edu.eg}
\affiliation{Physics Department, Faculty of Science, Ain Shams University, Cairo 11566, Egypt.}

\author{M. Medhat}
\email[]{mmedhat61@sci.asu.edu.eg}
\affiliation{Physics Department, Faculty of Science, Ain Shams University, Cairo 11566, Egypt.}

%\date{\today}

\begin{abstract}
We investigate a system of four nearest neighbour bidirectional coupled phase oscillators of dissimilar initial frequencies in a ring at the changeover into a synchronizing state. There are twenty four permutations upon assigning the initial frequencies to the oscillators. The local interaction between the adjacent coupled phase oscillators introduces details to the synchronization features. Therefore, for the four unalike local coupled oscillators, we classify all possible arrangements into three classes, where each class contains eight configurations. The synchronized state appears at the distinctive coupling when the oscillators transit to synchrony having noticeable characteristics for each class. Also, the unison behaviour emerges when a well-defined phase condition is developed. We utilize this conspicuous phase condition to obtain a mathematical expression predicting the distinguishing coupling for each class once the oscillators have a common frequency. The obtained expression is given in terms of the initial frequencies, at the minute the four local coupled phase oscillators attain the same frequency. The analytic formula of the critical coupling allows us to obtain expressions usable to determine the phase differences at the swop into a synchronization stage.
\end{abstract}


\maketitle

\section{Introduction}
The dynamics of systems in physics, chemistry, computing, biology, engineering, social relationships and environmental science are modelled by nonlinear coupled oscillators \cite{1,2,3,4,5,6}. The shared phenomenon among the previously mentioned wide spread areas of research, though they have dissimilar dynamics, is the synchronization of the mutually interacting oscillators \cite{7,8,9,10,11,12}. Thus coupled oscillators, under the effect of increasing coupling strength, operate in harmony. The Kuramoto model, a simple paradigmatic example of the non-identical coupled phase oscillators, is a successful and an efficient model that can describe the dynamics in many of the formerly stated topics of research. Hence, the Kuramoto model is capable to explain the synchronization phenomenon that has been noticed \cite{13,14,15,16,17,18,19,20}.

The original Kuramoto model represents all-to-all coupled phase oscillators of non-identical initial frequencies [1, 6]. However, in many cases, the nearest neighbour interactions between the dissimilar oscillators are necessary \cite{21,22,23,24}. This limitations to neighbouring interactions characterize the local coupled Kuramoto model (LCKM of coupled phase oscillators). The coupling between the nearest neighbour oscillators, in the LCKM, can be made unidirectional or bidirectional \cite{25, 26}. Here we are interested in the bidirectional version of the local coupled Kuramoto model (BLCKM) at the incipient of synchronization. The synchronization appears when the non-identical oscillators transit from a desynchronization feature to a unison behaviour at a threshold coupling value \cite{27, 28}. Thus, at the stage of a complete synchronization all the non-identical oscillators in the system become phase locked. Subsequently, the oscillators own a common frequency regardless the different natural frequencies of oscillators prior to coupling. In this case, the BLCKM neither analytically solved for a small number of oscillators ($N > 3$) nor for a large number of oscillators. Analytic solutions, for a few and a finite number of non-identical nearest neighbour coupled phase oscillators at synchronization, represent challenging problems. However, procurement analytic solutions is crucial from theoretical and practical points of view \cite{27,28,29,30,31,32}. Particularly, solutions for four and a few non-identical BLCKM oscillators are very important in order to recognise the synchronization mechanisms and the dynamics leading to the harmony behaviour. Understanding the dynamics of four disparate BLCKM oscillators will permit us realizing the differences between the four local coupled and the four global coupled systems. The case of four global coupled system is solved analytically at the full phase locking state \cite{33}. Moreover, the solutions for a few oscillators, are significant not only because these systems emerge in many applications but also necessary to guide us to find a method to solve the synchronization problem in a finite number of local coupled phase oscillators. Once we figure out cases of a few and a finite number of oscillators, we shall progress to be aware of systems of a very large number of oscillators.

Recently, the synchronization mechanism of a BLCKM of non-identical three oscillators has been analyzed and solved analytically \cite{34, 35} at the stage of a complete frequency synchronization. The situation of non-identical three oscillators represents a case of bidirectional coupled phase oscillators that allows to solve the unidirectional coupled three oscillators \cite{34}. However, the case of three BLCKM oscillators is a simple case where local and global couplings are equal. The problem of four BLCKM oscillators is different in comparison to the case of three BLCKM oscillators. We expect the dynamics, for four non-identical BLCKM oscillators, to be richer. The limitation of the interactions to be nearest neighbour, in this case of four non-identical BLCKM oscillators, introduces further details in order to obtain a solution. These details arise because we have twenty four different ways to arrange the initial frequencies of the nearest neighbour oscillators in a ring topology. Sequentially, the synchronization features of the adjacent oscillators are influenced.

In this work, we present a study of the four non-identical BLCKM oscillators at the instant of a perfect frequency synchronization. An analytic solution is obtained not only for the critical coupling constant but also for the phase differences between each two oscillators. The obtained expressions depend only on the natural frequencies of the oscillators.
This work is organized as follows: In section II, we present a model of four dissimilar BLCKM oscillators. In section III, we show the synchronization trees and the different patterns of the synchronizations of the non-identical four BLCKM oscillators. Also, in section III, we present the classification of the three different classes. A solution of the system of unalike BLCKM oscillators and hence an exact mathematical expression for the critical coupling constant is provided in section IV. In section V, we introduce furthest remarks on the decisive coupling at synchronization. We obtain, in section VI, formulas for the phase differences at the appearance of synchronization. A conclusion is given in section VII.

\section{Model of The Four Non-identical Local-Coupled Phase Oscillators}
The non-identical four BLCKM oscillators are symbolized mathematically by \cite{26,27,28}
\be
\dot{\theta}_i=\omega_i+ \frac{K}{3} [\sin(\theta_{i+1}-\theta_i) +
\sin(\theta_{i-1} - \theta_i)],
\label{eq:one}
\ee
for $i=1,2,3,4$. According to model (1), $\omega_i$ and $\theta_i(t)$ are the initial frequency and the instantaneous phase of the $i^{th}$ oscillator, successively, while $\dot\theta_i$ is its instantaneous frequency. The quantities $(\theta_{i+1}-\theta_i)$ and $(\theta_{i-1}-\theta_i)$ in (1) are phase differences. The coupling constant is $K$. The oscillators are in a ring, where periodic boundary conditions are applied (if $i = 4$ then $i + 1 = 1$ and when $i = 1$ thence
$i - 1 = 4$). In general, using $(1/4)\sum_{j=1}^4 \omega_j=\omega_o=0$, $i = 1, 2, 3, 4$, does not alter the features of system (1). At the moment a complete frequency synchronization stage appears, when $K = K_c$ ($K_c$ is the critical coupling), the frequencies of all oscillators become $\omega_o = 0$ and all the time evolution of phases $\dot\theta_i(t)=0$, $i = 1, 2, 3, 4$. In addition, the quantities $\dot\theta_i$, the phase differences $(\theta_{i+1}-\theta_i)$ and $(\theta_{i-1}-\theta_i)$ as well as their time evolution $(\dot\theta_{i+1}-\dot\theta_i)$ and $(\dot\theta_{i-1}-\dot\theta_i)$, for $i = 1, 2, 3, 4$, turn into time independent quantities at $K_c$. Thus, a phase lock is established once a stage of complete frequency synchronization at $K_c$ appears. The phase lock is defined by a prime phase difference $\phi_j=(\theta_j-\theta_k)$, $j \neq k$, which becomes $\pi/2$ (phase lock condition) \cite{26,27,28}.
The time evolution of the frequency differences according to (1) are given by
\be
\dot \phi_i = \Delta_i + \frac{K}{3} [\sin(\theta_{i+2}-\theta_i+1) - 2\sin (\theta_{i+1}-\theta_i) + \sin (\theta_i-\theta_{i-1})],
\label{eq:two}
\ee
for $i=1,2,3,4$. In equation (2), the time progression of the phase differences is $\dot\phi_i=(\dot\theta_{i+1}-\dot\theta_i)$, $i = 1, 2, 3, 4$. The quantities $\Delta_i=(\omega_{i+1}-\omega_i)$, $i = 1, 2, 3, 4$, are always the frequency differences between two successive pairs of oscillators. According to equation (2), the frequency synchronization of the four dissimilar BLCKM oscillators occurs when all the time evolution of the frequency differences $\dot\phi_i=0$, and one of the phase differences $\phi_j=\pi/2$. This phase lock condition corresponds to the absolute maximum frequency difference $\Delta_j$. We shall define exactly $\phi_j=\pi/2$ and $\Delta_j$ tardier in the text, for the four BLCKM oscillators, when we introduce the details of the twenty four configurations of the assigning the initial frequencies of the four oscillators over the set \{$\omega_i$\}.

The conditions considered in this work for the unalike oscillators of BLCKM, when allocating the initial frequencies, are restricted to not allowing any two oscillators or more to have the same preliminary frequencies. Precisely, the frequency differences between any two close oscillators must be $\geq 0.01$ in addition to the condition $(1/4)\sum_{j=1}^4 \omega_j=\omega_o=0$, $i = 1, 2, 3, 4$.

\section{Synchronization Trees, Classifications of Configurations and Classes}
The different choices of the natural frequencies set \{$\omega_i$\}, $i=1,2,3,4$, for four dissimilar oscillators, are twenty four. Thus, there are twenty four different permutations in order to distribute the frequencies over the orderly arranged set \{$\omega_{max},\omega_{m>},\omega_{m<},\omega_{min}$\}. The maximum and the minimum values of the initial frequencies are $\omega_{max}$ and $\omega_{min}$. The other two initial frequencies, in between the maximum and the minimum frequencies, are $\omega_{m>}$ and $\omega_{m<}$. Always, in any arrangement, we find \{$\omega_{max}>\omega_{m>}>\omega_{m<}>\omega_{min}$\}. Accordingly, the frequency differences have to be determined as the differences between the initial frequencies of the nearest neighbour oscillators only. Thus, the four incongruent oscillators of BLCKM will have phase locked states depending on the prearrangement of the initial frequencies.

Motivated by the solution of the four all-to-all coupled phase oscillators \cite{33}, we expect that the absolute maximum frequency difference (between the maximum and the minimum frequencies) as well as the frequency difference (between the other two intermediate frequencies that allocate between the maximum and minimum frequencies) are the effective parameters to control the phase locked state. Also, the maximum and minimum frequencies contribute to ruling the sync state in addition to the two previously stated frequency differences. Unlike the global case, the interactions in the local and bidirectional coupled model are reduced to the nearest neighbour oscillators merely. Therefore, we anticipate that the previously mentioned parameters affect the synchronizing state rely on the different permutations of the four dissimilar BLCKM oscillators (as will be explained well ahead). Consequently, the formerly mentioned parameters must be given in terms of the initial frequencies of the non-identical nearest neighbour oscillators. However, the definitions of the formerly specified parameters are not unique for the all twenty four different arrangements.

Accordingly, as we regard nearest neighbour interactions, we distinguish three groups of arrangements of initial frequencies to be distributed over \{$\omega_{max},\omega_{m>},\omega_{m<},\omega_{min}$\}. Each group is composed of eight configurations. We shall rename the three groups as class I, class II and class III. For each class, we define the nearest neighbour frequency difference $\tilde{\Delta}_{max}=|\Delta_i^{nn}|_{max}$, for $i = 1, 2, 3, 4$, where $nn$ refers to nearest neighbour, which comes as the absolute maximum difference between two frequencies of two nearest neighbour oscillators. We identify another nearest neighbour frequency difference $\tilde{\Delta}_{m}=(\Delta_i^{nn})_{m}$ as obtained by means of the difference between the other two nearest neighbour oscillators’s frequencies that are not involved in defining $\tilde{\Delta}_{max}=|\Delta_i^{nn}|_{max}$. Therefore, we have distinct characterizations of these two quantities $\tilde{\Delta}_{max}>0$ and $\tilde{\Delta}_{m}$ for each class. We shall explore, in details later within the text, the different configurations as well as the definitions of both quantities $\tilde{\Delta}_{max}>0$ and $\tilde{\Delta}_{m}$ within each class. Also, we shall introduce the rearranged frequencies $\tilde{\omega}_{max}$, $\tilde{\omega}_{m>}$, $\tilde{\omega}_{m<}$ and $\tilde{\omega}_{min}$ as well as the relocated phases $\tilde{\theta}_{max}$, $\tilde{\theta}_{m>}$, $\tilde{\theta}_{m<}$ and $\tilde{\theta}_{min}$. For each class, these reordered quantities $(\tilde{\omega}_{max}$, $\tilde{\omega}_{m>}$, $\tilde{\omega}_{m<}$ and $\tilde{\omega}_{min})$ are necessary for redefining the nearest neighbour frequency differences. Consequently, we shall utilize all the quantities as defined above to rewrite system (1) in a unified form for all classes taking into considerations the reordered phases. This is a required step in order to obtain a solution for system (1), as it will be clear afterward in the text. Specifically, we look for a solution that expresses exactly the critical coupling as a function of nearest neighbouring frequency differences $\tilde{\Delta}_{max}$ and $\tilde{\Delta}_{m}$ in addition to nearest neighbour frequencies $\tilde{\omega}_{max}$ and $\tilde{\omega}_{min}$.

\subsection{Class I}
\begin{figure}
\includegraphics[scale=0.5]{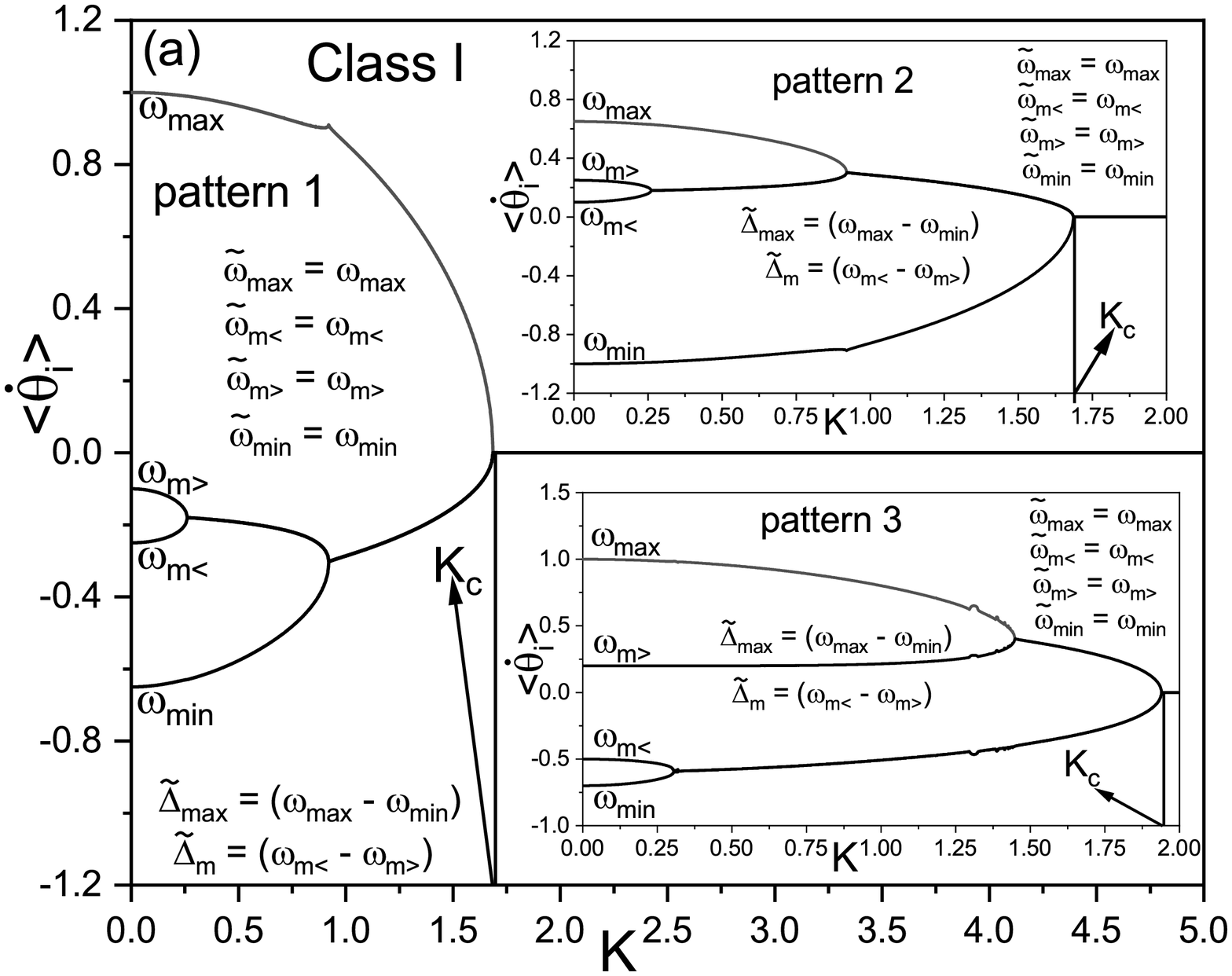}
\includegraphics[scale=0.5]{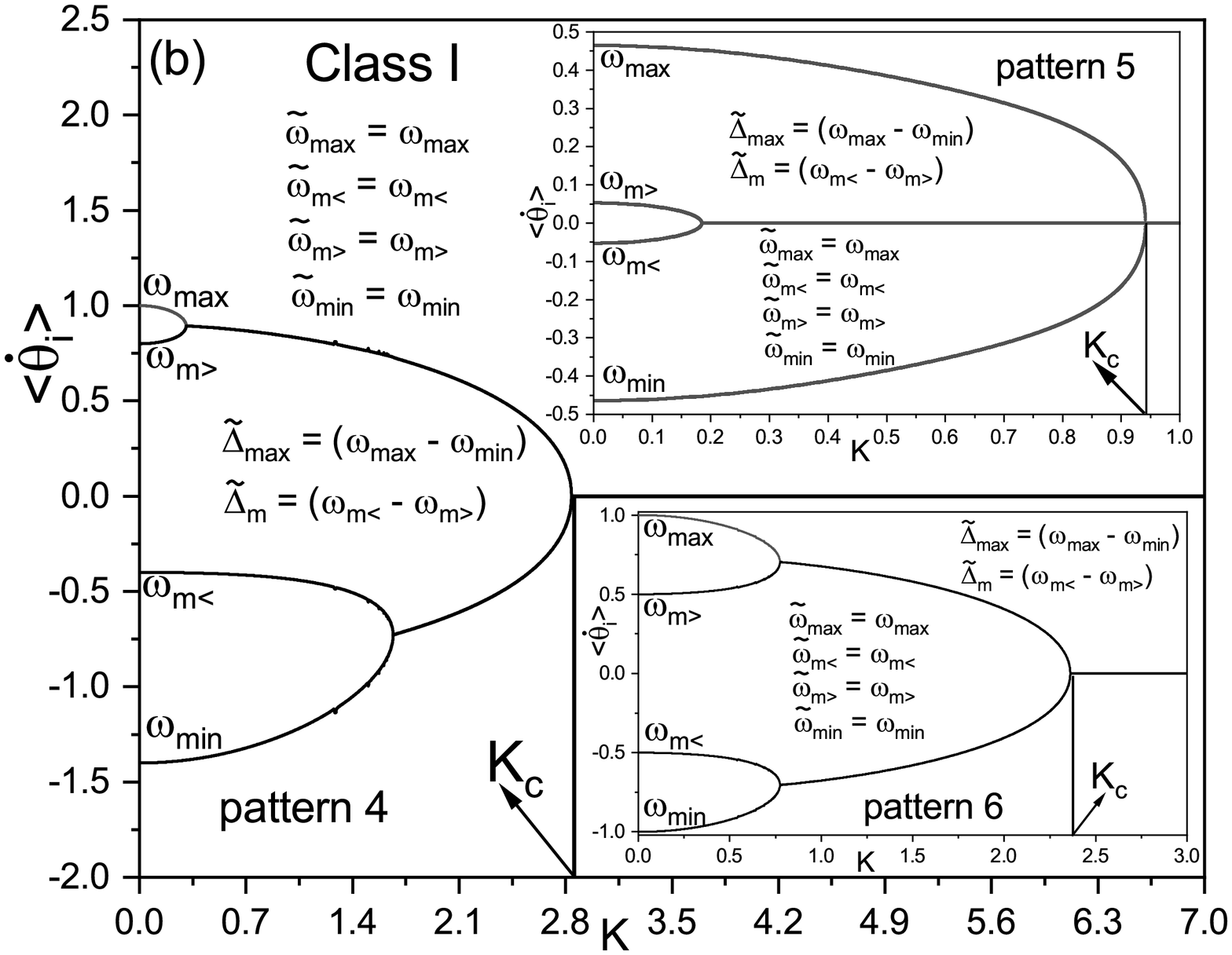}
\caption{\label{fig:one} The examples of the synchronization trees of the dissimilar four BLCKM oscillators, for class I. Conditions to obtain different patterns are indicated in the legends of each diagram. (a) The main plot shows $1^{st}$ pattern. The upper inset diagram demonstrates $2_{nd}$ pattern. The lower inset graph makes obvious $3^{rd}$ pattern. (b) The main plot presents $4^{th}$ pattern. The upper inset reveals $5^{th}$ pattern while the lower inset illustrates $6^{th}$ pattern. }
\end{figure}
Class I appears when each consecutive pair of initial frequencies $(\tilde{\omega}_{max}$ $\&$ $\tilde{\omega}_{m>})$, $(\tilde{\omega}_{m>}$ $\&$ $\tilde{\omega}_{m<})$, and $(\tilde{\omega}_{m<}$ $\&$ $\tilde{\omega}_{min})$ are possessed by two nearest neighbour oscillators. In addition, the pair $(\tilde{\omega}_{max}$ $\&$ $\tilde{\omega}_{min})$ is composed of nearest neighbours because of the ring topology.

The plots of Figure 1(a, b) depicts examples of the synchronization trees, for class I. The main plot of Figure 1a (pattern 1) appears, at the sync state, when one oscillator of $\omega_{max}$ alone encounters a cluster of three oscillators of $\omega_{m>}$, $\omega_{m<}$ and $\omega_{min}$. The upper inset graph of Figure 1a (pattern 2), illustrates the synchronization as soon as the three oscillators’ cluster of $\omega_{max}$, $\omega_{m>}$ and $\omega_{m<}$ joins the other oscillator of $\omega_{min}$, at a moment a synchronization stage comes out. The lower insert diagram of Figure 1 (pattern 3) elucidates the synchronization, once a cluster of two oscillators of $\omega_{max}$ and $\omega_{m>}$ connects to a cluster of the other two oscillators of $\omega_{m<}$ and $\omega_{min}$ at $K_c$, where the gap between $\omega_{max}$ and $\omega_{m>}$ is larger than that of $\omega_{m<}$ and $\omega_{min}$. The main plot of Figure 1b (pattern 4) points out to the synchronization once the oscillators are ordered to be a cluster of two oscillators of original frequencies $\omega_{max}$ and $\omega_{m>}$ attaching to a cluster of the other two oscillators of the earliest frequencies $\omega_{m<}$ and $\omega_{min}$ at $K_c$ but the separation between $\omega_{max}$ and $\omega_{m>}$ is smaller than the splitting between $\omega_{m<}$ and $\omega_{min}$. The upper inset of Figure 1b (pattern 5) shows the synchronization at $K_c$ between a cluster of $\omega_{m<}$ and $\omega_{m>}$ when it meets a cluster of $\omega_{max}$ and $\omega_{min}$. The diagram in the lower inset of Figure 1b shows the pattern 6 when a group of $\omega_{m>}$ and $\omega_{m<}$ joins another of $\omega_{max}$ and $\omega_{min}$. In the patterns 4 and 5 both clusters have the same widths. The difference between patterns 5 and 6 appears due to the separation between the middle oscillators in the $5^{th}$ is smaller than that in the $6^{th}$. We find always, in class I, the phase difference satisfying the phase lock condition $\phi_{max}=(\theta_{max}-\theta_{min})=\pi/2$ at $K_c$ is corresponding to the frequency difference $\tilde{\Delta}_{max}>0$.
\begin{table}
\caption{The eight assemblies that define class I. The quantities $\tilde{\Delta}_{max}>0$ and $\tilde{\Delta}_{m}<0$ are given for each specific configuration.}
%{tabular environment}%
\label{table-one}
\begin{tabular}{|c|c|c|}
\hline
 configuration & $\tilde{\Delta}_{max}>0$ & $\tilde{\Delta}_{m}<0$  \\
$(\omega_{\max},\omega_{m>},\omega_{m<},\omega_{min})$  & $\tilde{\Delta}_{max}=(\tilde{\omega}_{max}-\tilde{\omega}_{min})$ &  $\tilde{\Delta}_{m}=(\tilde{\omega}_{m<}-\tilde{\omega}_{m>})$  \\
\hline
 $(\omega_1,\omega_2,\omega_3,\omega_4)$  & $\tilde{\Delta}_{max}=(\omega_1-\omega_4)$  & $\tilde{\Delta}_{m}=(\omega_3-\omega_2)$  \\
\hline
 $(\omega_1,\omega_4,\omega_3,\omega_2)$  & $\tilde{\Delta}_{max}=(\omega_1-\omega_2)$   & $\tilde{\Delta}_{m}=(\omega_3-\omega_4)$ \\
\hline
 $(\omega_2,\omega_3,\omega_4,\omega_1)$  & $\tilde{\Delta}_{max}=(\omega_2-\omega_1)$  & $\tilde{\Delta}_{m}=(\omega_4-\omega_3)$  \\
\hline
 $(\omega_2,\omega_1,\omega_4,\omega_3)$  & $\tilde{\Delta}_{max}=(\omega_2-\omega_3)$  & $\tilde{\Delta}_{m}=(\omega_4-\omega_1)$  \\
\hline
 $(\omega_3,\omega_4,\omega_1,\omega_2)$  & $\tilde{\Delta}_{max}=(\omega_3-\omega_2)$  & $\tilde{\Delta}_{m}=(\omega_1-\omega_4)$    \\
\hline
 $(\omega_3,\omega_2,\omega_1,\omega_4)$  & $\tilde{\Delta}_{max}=(\omega_3-\omega_4)$  & $\tilde{\Delta}_{m}=(\omega_1-\omega_2)$ \\
\hline
 $(\omega_4,\omega_1,\omega_2,\omega_3)$  & $\tilde{\Delta}_{max}=(\omega_4-\omega_3)$  & $\tilde{\Delta}_{m}=(\omega_2-\omega_1)$ \\
\hline
 $(\omega_4,\omega_3,\omega_2,\omega_1)$  & $\tilde{\Delta}_{max}=(\omega_4-\omega_1)$  & $\tilde{\Delta}_{m}=(\omega_2-\omega_3)$ \\
 \hline
\end{tabular}
\end{table}
As clearly appears in Figure 1, class I is obtained when all oscillators appear nearest neighbours as well as they possess $\tilde{\omega}_{max}=\omega_{max}$, $\tilde{\omega}_{min}=\omega_{min}$, $\tilde{\omega}_{m>}=\omega_{m>}$, and $\tilde{\omega}_{m<}=\omega_{m<}$. Thus, class I is distinguished by $|\Delta_i^{nn}|_{max}\equiv\tilde{\Delta}_{max}=(\omega_{max}-\omega_{min})>0$ and $(\Delta_i^{nn})_{m}\equiv \tilde{\Delta}_m=(\omega_{m<}-\omega_{m>})<0$. Also, the phases are $\tilde{\theta}_{max}=\theta_{max}$, $\tilde{\theta}_{m>}=\theta_{m>}$, $\tilde{\theta}_{m<}=\theta_{m<}$ and $\tilde{\theta}_{min}=\theta_{min}$. The eight arrangements of oscillators corresponding to class I are summarized in table I. As indicated in table I, the left column shows the different eight configurations to distribute the initial frequencies on the set \{$\omega_{max}>\omega_{m>}>\omega_{m<}>\omega_{min}$\}. The other two columns to the right present the quantities $\tilde{\Delta}_{max}>0$ and $\tilde{\Delta}_m<0$, respectively.

\subsection{Class II}
\begin{figure}
\includegraphics[scale=0.5]{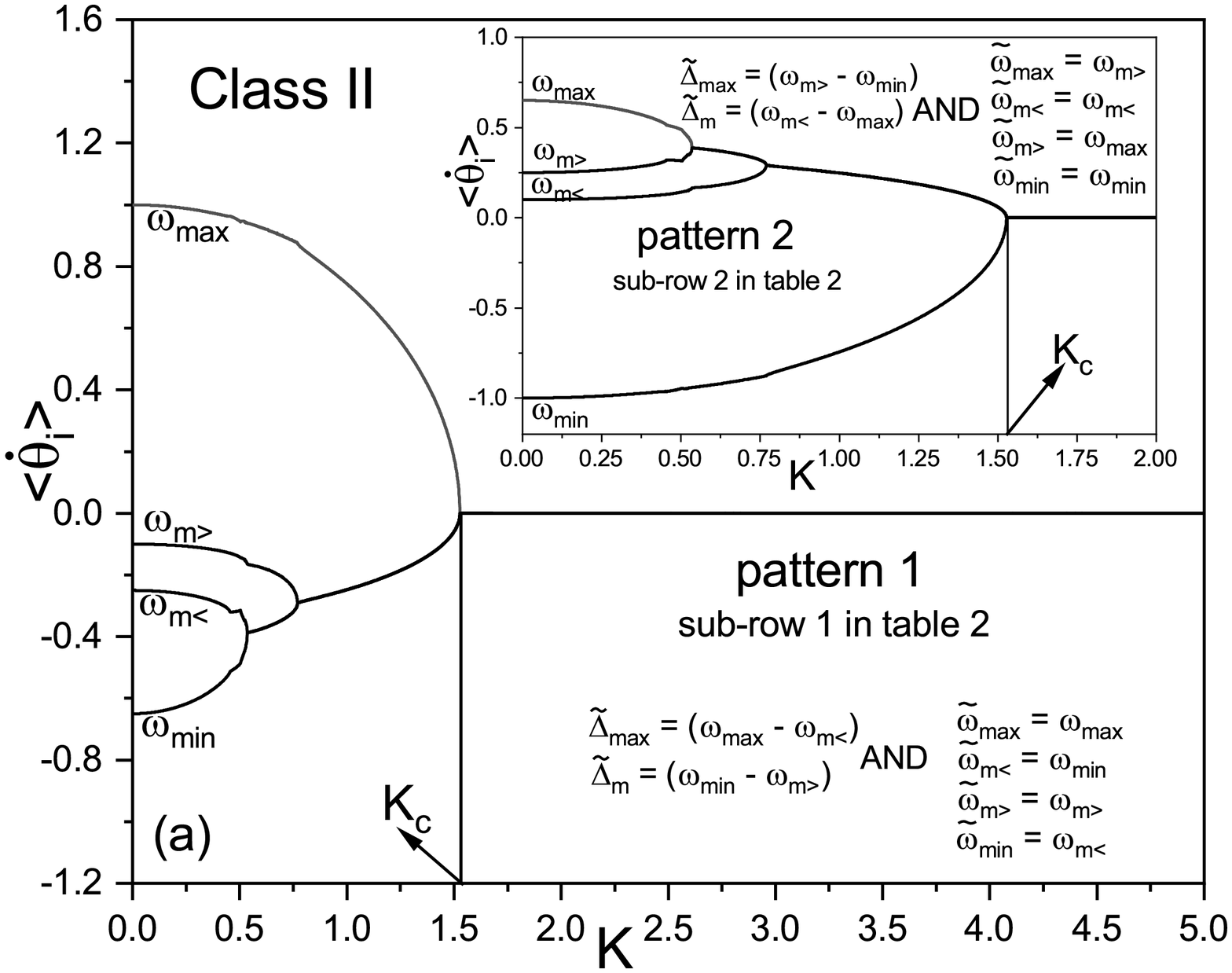}
\includegraphics[scale=0.5]{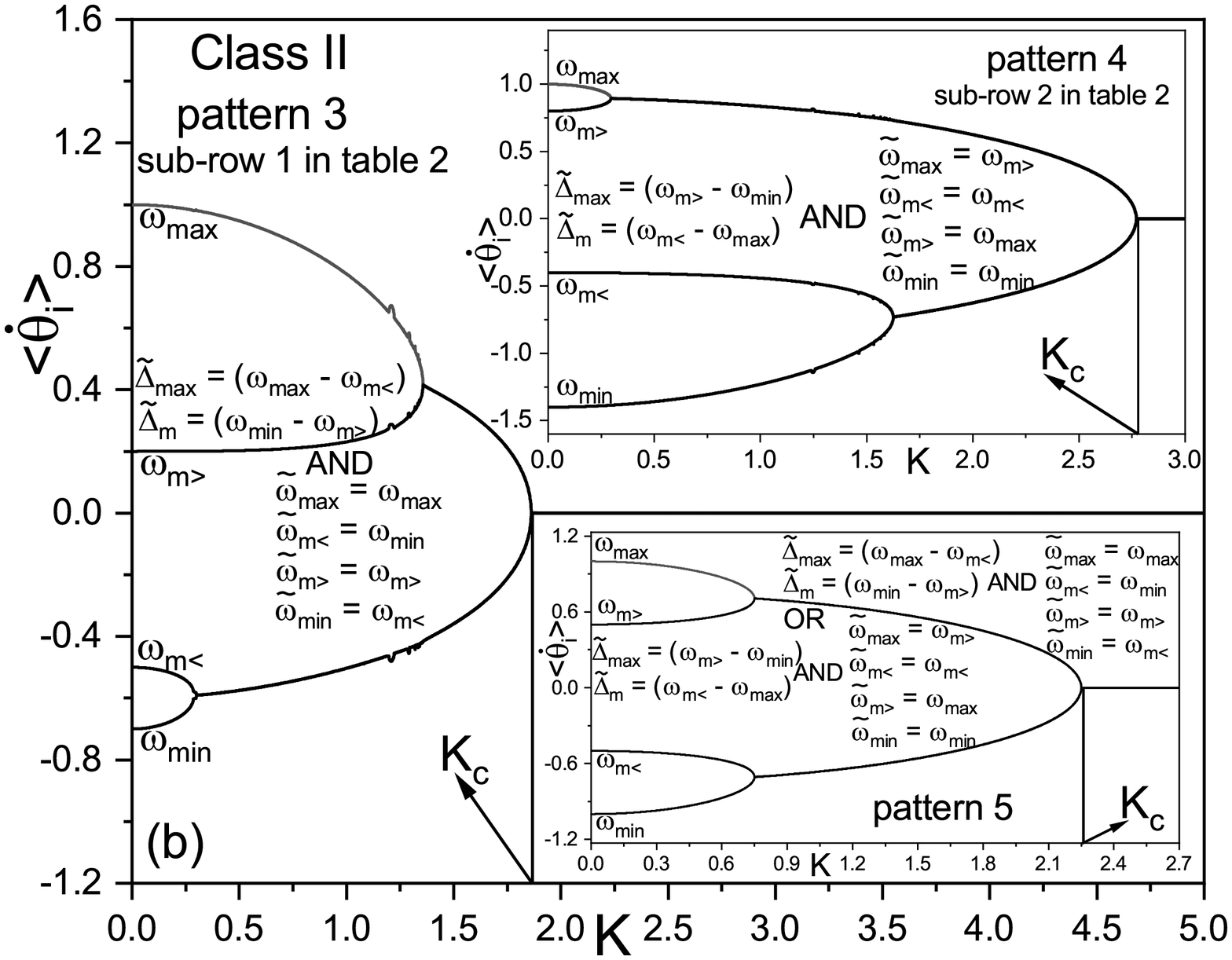}
\caption{\label{fig:two} Samples of the synchronization hierarchies of four BLCKM oscillators, for class II, when the coupling constant increases for distinctive configurations of \{$\omega_{\max}>\omega_{m>}>\omega_{m<}>\omega_{min}$\}. Settings to find different patterns are specified in the legends of each graph. (a) The principal drawing points to the pattern 1. The higher insert figure puts on view the pattern 2. (b) The main diagram illustrates the pattern 3 and the upper inset plot indicates the pattern 4 while the lower inset shows the pattern 5. The legends of all plots show the two condition of the choices (as explained in table II). In all plots, we refer to table II, where each row is composed of two sub-rows (first and second).}
\end{figure}
Class II exists as neither the two oscillators of frequencies $(\omega_{max} \& \omega_{min})$ nor the couple oscillators of frequencies $(\omega_{m>} \& \omega_{m<})$ are handled nearest neighbours. The pairing oscillators of frequencies $(\omega_{max} \& \omega_{m>})$ are nearest neighbours as well as the combining oscillators of frequencies $(\omega_{m<} \& \omega_{min})$ are nearest neighbours too. The major drawing of Figure 2a, at the unison state, presents a pattern 1 of the synchronization feature once one oscillator of $\omega_{max}$ comes together with the group of the three oscillators of $\omega_{m>}$, $\omega_{m<}$ and $\omega_{min}$. The upper inset of Figure 2a, indicates the synchronization behaviour (pattern 2) as soon as a cluster of three oscillators of $\omega_{max}$, $\omega_{m>}$ and $\omega_{m<}$ meets the other oscillator of $\omega_{min}$ at the critical coupling. The main graph of Figure 2b, displays the synchronization tree (pattern 3) as a group of two oscillators of $\omega_{max}$ and $\omega_{m>}$ assembles another of $\omega_{m<}$ and $\omega_{min}$ at $K_c$, where the interval between $\omega_{max}$ and $\omega_{m>}$ is larger than the difference between $\omega_{m<}$ and $\omega_{min}$. The diagram in the lower inset of Figure 2b (pattern 4) shows the cluster of $\omega_{max}$ and $\omega_{m>}$ once it joins the other one of $\omega_{m<}$ and $\omega_{min}$ at the critical coupling, when the space between $\omega_{max}$ and $\omega_{m>}$ is smaller than the split between $\omega_{m<}$ and $\omega_{min}$. Unlike class I, we cannot find in class II a partial synchronization between the middle oscillators because they are not nearest neighbours. As a result, for class II, the nearest neighbour frequency difference $|\Delta_i^{nn}|_{max} \equiv \tilde{\Delta}_{max} \neq (\omega_{max}-\omega_{min})$. In addition, the other nearest neighbour frequency difference $(\Delta_i^{nn})_{m} \equiv \tilde{\Delta}_m \neq (\omega_{m<}-\omega_{m>})$. Therefore, the redefined frequencies $\tilde{\omega}_{max}$, $\tilde{\omega}_{min}$, $\tilde{\omega}_{m>}$ and $\tilde{\omega}_{m<}$, as well as their consistent rearranged phases $\tilde{\theta}_{max}$, $\tilde{\theta}_{min}$, $\tilde{\theta}_{m>}$ and $\tilde{\theta}_{m<}$ will be different than the case of class I. We shall delineate all the previously mentioned quantities.
%\squeezetable
%\begingroup
%\begin{singlespace}
\begin{table}[t]
\caption{The eight arrangements of class II. The frequency differences $\tilde{\Delta}_{max}>0$ and $\tilde{\Delta}_{m}<0$ are specified for each particular configuration. Each row has two sub-rows in accord to the resettled nearest neighbour frequencies as well as the relation between $\tilde{\Delta}_{max}$ and $|\tilde{\Delta}_{m}|$. The reallocated phases are defined.}
%{tabular environment}%
\label{table-two}
\begin{tabular}{|c|c|c|c|}
\hline
 configuration & $\tilde{\Delta}_{max}>0$ & $\tilde{\Delta}_{m}<0$ &  Phases \\
$(\omega_{\max},\omega_{m>},\omega_{m<},\omega_{min})$  & $\tilde{\Delta}_{max}=(\tilde{\omega}_{max}-\tilde{\omega}_{min})$ &  $\tilde{\Delta}_{m}=(\tilde{\omega}_{m<}-\tilde{\omega}_{m>})$ &    \\
%\hline
\hline
 $(\omega_1,\omega_2,\omega_4,\omega_3)$  & $(\omega_1-\omega_4)>(\omega_2-\omega_3)$  & $\tilde{\omega}_{m>}=\omega_2$ $\&$ $\tilde{\omega}_{m<}=\omega_3$ & $\tilde{\theta}_{max}=\theta_1$ $\&$ $\tilde{\theta}_{m>}=\theta_2$ \\
  $\tilde{\omega}_{max}=\omega_1$ $\&$ $\tilde{\omega}_{min}=\omega_4$ & $\tilde{\Delta}_{max}=(\omega_1-\omega_4)$ & $\tilde{\Delta}_{m}=(\omega_3-\omega_2)$ & $\tilde{\theta}_{min}=\theta_4$ $\&$ $\tilde{\theta}_{m<}=\theta_3$ \\
\hline
%\hdashline
                                          & $(\omega_2-\omega_3)>(\omega_1-\omega_4)$ & $\tilde{\omega}_{m>}=\omega_1$ $\&$ $\tilde{\omega}_{m<}=\omega_4$ & $\tilde{\theta}_{max}=\theta_2$ $\&$ $\tilde{\theta}_{m>}=\theta_1$ \\
 $\tilde{\omega}_{max}=\omega_2$ $\&$ $\tilde{\omega}_{min}=\omega_3$ & $\tilde{\Delta}_{max}=(\omega_2-\omega_3)$ & $\tilde{\Delta}_{m}=(\omega_4-\omega_1)$ & $\tilde{\theta}_{min}=\theta_3$ $\&$ $\tilde{\theta}_{m>}=\theta_4$ \\
%\hline
\hline
 $(\omega_1,\omega_4,\omega_2,\omega_3)$  & $(\omega_1-\omega_2)>(\omega_4-\omega_3)$  & $\tilde{\omega}_{m>}=\omega_4$ $\&$ $\tilde{\omega}_{m<}=\omega_3$ & $\tilde{\theta}_{max}=\theta_1$ $\&$ $\tilde{\theta}_{m>}=\theta_4$ \\
  $\tilde{\omega}_{max}=\omega_1$ $\&$ $\tilde{\omega}_{min}=\omega_2$ & $\tilde{\Delta}_{max}=(\omega_1-\omega_2)$ & $\tilde{\Delta}_{m}=(\omega_3-\omega_4)$ & $\tilde{\theta}_{min}=\theta_2$ $\&$ $\tilde{\theta}_{m<}=\theta_3$ \\
\hline
%\hdashline
                                          & $(\omega_4-\omega_3)>(\omega_1-\omega_2)$ & $\tilde{\omega}_{m>}=\omega_1$ $\&$ $\tilde{\omega}_{m<}=\omega_2$ & $\tilde{\theta}_{max}=\theta_4$ $\&$ $\tilde{\theta}_{m>}=\theta_1$ \\
 $\tilde{\omega}_{max}=\omega_4$ $\&$ $\tilde{\omega}_{min}=\omega_3$ & $\tilde{\Delta}_{max}=(\omega_4-\omega_3)$ & $\tilde{\Delta}_{m}=(\omega_2-\omega_1)$ & $\tilde{\theta}_{min}=\theta_3$ $\&$ $\tilde{\theta}_{m>}=\theta_2$ \\
%\hline
\hline
 $(\omega_2,\omega_3,\omega_1,\omega_4)$  & $(\omega_2-\omega_1)>(\omega_3-\omega_4)$  & $\tilde{\omega}_{m>}=\omega_3$ $\&$ $\tilde{\omega}_{m<}=\omega_4$ & $\tilde{\theta}_{max}=\theta_2$ $\&$ $\tilde{\theta}_{m>}=\theta_3$ \\
  $\tilde{\omega}_{max}=\omega_2$ $\&$ $\tilde{\omega}_{min}=\omega_1$ & $\tilde{\Delta}_{max}=(\omega_2-\omega_1)$ & $\tilde{\Delta}_{m}=(\omega_4-\omega_3)$ & $\tilde{\theta}_{min}=\theta_1$ $\&$ $\tilde{\theta}_{m<}=\theta_4$ \\
\hline
%\hdashline
                                          & $(\omega_3-\omega_4)>(\omega_2-\omega_1)$ & $\tilde{\omega}_{m>}=\omega_1$ $\&$ $\tilde{\omega}_{m<}=\omega_2$ & $\tilde{\theta}_{max}=\theta_3$ $\&$ $\tilde{\theta}_{m>}=\theta_1$ \\
 $\tilde{\omega}_{max}=\omega_3$ $\&$ $\tilde{\omega}_{min}=\omega_4$ & $\tilde{\Delta}_{max}=(\omega_3-\omega_4)$ & $\tilde{\Delta}_{m}=(\omega_2-\omega_1)$ & $\tilde{\theta}_{min}=\theta_4$ $\&$ $\tilde{\theta}_{m>}=\theta_2$ \\
%\hline
\hline
 $(\omega_2,\omega_1,\omega_3,\omega_4)$  & $(\omega_2-\omega_3)>(\omega_1-\omega_4)$  & $\tilde{\omega}_{m>}=\omega_1$ $\&$ $\tilde{\omega}_{m<}=\omega_4$ & $\tilde{\theta}_{max}=\theta_2$ $\&$ $\tilde{\theta}_{m>}=\theta_1$ \\
  $\tilde{\omega}_{max}=\omega_2$ $\&$ $\tilde{\omega}_{min}=\omega_3$ & $\tilde{\Delta}_{max}=(\omega_2-\omega_3)$ & $\tilde{\Delta}_{m}=(\omega_4-\omega_1)$ & $\tilde{\theta}_{min}=\theta_3$ $\&$ $\tilde{\theta}_{m<}=\theta_4$ \\
\hline
                                          & $(\omega_1-\omega_4)>(\omega_2-\omega_3)$ & $\tilde{\omega}_{m>}=\omega_2$ $\&$ $\tilde{\omega}_{m<}=\omega_3$ & $\tilde{\theta}_{max}=\theta_1$ $\&$ $\tilde{\theta}_{m>}=\theta_2$ \\
 $\tilde{\omega}_{max}=\omega_1$ $\&$ $\tilde{\omega}_{min}=\omega_4$ & $\tilde{\Delta}_{max}=(\omega_1-\omega_4)$ & $\tilde{\Delta}_{m}=(\omega_3-\omega_2)$ & $\tilde{\theta}_{min}=\theta_4$ $\&$ $\tilde{\theta}_{m>}=\theta_3$ \\
%\hline
\hline
 $(\omega_3,\omega_4,\omega_2,\omega_1)$  & $(\omega_3-\omega_2)>(\omega_4-\omega_1)$  & $\tilde{\omega}_{m>}=\omega_2$ $\&$ $\tilde{\omega}_{m<}=\omega_3$ & $\tilde{\theta}_{max}=\theta_3$ $\&$ $\tilde{\theta}_{m>}=\theta_4$ \\
  $\tilde{\omega}_{max}=\omega_3$ $\&$ $\tilde{\omega}_{min}=\omega_2$ & $\tilde{\Delta}_{max}=(\omega_3-\omega_2)$ & $\tilde{\Delta}_{m}=(\omega_1-\omega_4)$ & $\tilde{\theta}_{min}=\theta_2$ $\&$ $\tilde{\theta}_{m<}=\theta_1$ \\
\hline
                                          & $(\omega_4-\omega_1)>(\omega_3-\omega_2)$ & $\tilde{\omega}_{m>}=\omega_2$ $\&$ $\tilde{\omega}_{m<}=\omega_3$ & $\tilde{\theta}_{max}=\theta_4$ $\&$ $\tilde{\theta}_{m>}=\theta_3$ \\
 $\tilde{\omega}_{max}=\omega_4$ $\&$ $\tilde{\omega}_{min}=\omega_1$ & $\tilde{\Delta}_{max}=(\omega_4-\omega_1)$ & $\tilde{\Delta}_{m}=(\omega_2-\omega_3)$ & $\tilde{\theta}_{min}=\theta_1$ $\&$ $\tilde{\theta}_{m>}=\theta_2$ \\
%\hline
\hline
 $(\omega_3,\omega_2,\omega_4,\omega_1)$  & $(\omega_3-\omega_4)>(\omega_2-\omega_1)$  & $\tilde{\omega}_{m>}=\omega_2$ $\&$ $\tilde{\omega}_{m<}=\omega_1$ & $\tilde{\theta}_{max}=\theta_3$ $\&$ $\tilde{\theta}_{m>}=\theta_2$ \\
  $\tilde{\omega}_{max}=\omega_3$ $\&$ $\tilde{\omega}_{min}=\omega_4$ & $\tilde{\Delta}_{max}=(\omega_3-\omega_4)$ & $\tilde{\Delta}_{m}=(\omega_1-\omega_2)$ & $\tilde{\theta}_{min}=\theta_4$ $\&$ $\tilde{\theta}_{m<}=\theta_1$ \\
\hline
                                          & $(\omega_2-\omega_1)>(\omega_3-\omega_4)$ & $\tilde{\omega}_{m>}=\omega_3$ $\&$ $\tilde{\omega}_{m<}=\omega_4$ & $\tilde{\theta}_{max}=\theta_2$ $\&$ $\tilde{\theta}_{m>}=\theta_3$ \\
 $\tilde{\omega}_{max}=\omega_2$ $\&$ $\tilde{\omega}_{min}=\omega_1$ & $\tilde{\Delta}_{max}=(\omega_2-\omega_1)$ & $\tilde{\Delta}_{m}=(\omega_4-\omega_3)$ & $\tilde{\theta}_{min}=\theta_1$ $\&$ $\tilde{\theta}_{m>}=\theta_4$ \\
%\hline
\hline
 $(\omega_4,\omega_1,\omega_3,\omega_2)$  & $(\omega_4-\omega_3)>(\omega_1-\omega_2)$  & $\tilde{\omega}_{m>}=\omega_1$ $\&$ $\tilde{\omega}_{m<}=\omega_2$ & $\tilde{\theta}_{max}=\theta_4$ $\&$ $\tilde{\theta}_{m>}=\theta_1$ \\
  $\tilde{\omega}_{max}=\omega_4$ $\&$ $\tilde{\omega}_{min}=\omega_3$ & $\tilde{\Delta}_{max}=(\omega_4-\omega_3)$ & $\tilde{\Delta}_{m}=(\omega_2-\omega_1)$ & $\tilde{\theta}_{min}=\theta_3$ $\&$ $\tilde{\theta}_{m<}=\theta_2$ \\
\hline
                                          & $(\omega_1-\omega_2)>(\omega_4-\omega_3)$ & $\tilde{\omega}_{m>}=\omega_4$ $\&$ $\tilde{\omega}_{m<}=\omega_3$ & $\tilde{\theta}_{max}=\theta_1$ $\&$ $\tilde{\theta}_{m>}=\theta_4$ \\
 $\tilde{\omega}_{max}=\omega_1$ $\&$ $\tilde{\omega}_{min}=\omega_2$ & $\tilde{\Delta}_{max}=(\omega_1-\omega_2)$ & $\tilde{\Delta}_{m}=(\omega_3-\omega_4)$ & $\tilde{\theta}_{min}=\theta_2$ $\&$ $\tilde{\theta}_{m>}=\theta_3$ \\
%\hline
\hline
 $(\omega_4,\omega_3,\omega_1,\omega_2)$  & $(\omega_4-\omega_1)>(\omega_3-\omega_2)$  & $\tilde{\omega}_{m>}=\omega_3$ $\&$ $\tilde{\omega}_{m<}=\omega_2$ & $\tilde{\theta}_{max}=\theta_4$ $\&$ $\tilde{\theta}_{m>}=\theta_3$ \\
  $\tilde{\omega}_{max}=\omega_4$ $\&$ $\tilde{\omega}_{min}=\omega_1$ & $\tilde{\Delta}_{max}=(\omega_4-\omega_1)$ & $\tilde{\Delta}_{m}=(\omega_2-\omega_3)$ & $\tilde{\theta}_{min}=\theta_1$ $\&$ $\tilde{\theta}_{m<}=\theta_2$ \\
\hline
                                          & $(\omega_3-\omega_2)>(\omega_4-\omega_1)$ & $\tilde{\omega}_{m>}=\omega_4$ $\&$ $\tilde{\omega}_{m<}=\omega_1$ & $\tilde{\theta}_{max}=\theta_3$ $\&$ $\tilde{\theta}_{m>}=\theta_4$ \\
 $\tilde{\omega}_{max}=\omega_3$ $\&$ $\tilde{\omega}_{min}=\omega_2$ & $\tilde{\Delta}_{max}=(\omega_3-\omega_2)$ & $\tilde{\Delta}_{m}=(\omega_1-\omega_4)$ & $\tilde{\theta}_{min}=\theta_2$ $\&$ $\tilde{\theta}_{m>}=\theta_1$ \\
\hline
\end{tabular}
%\end{minipage}
%\end{center}
\end{table}
%\end{singlespace}
%\endgroup
As clearly seen in table II, the left column shows the different eight arrangements to put forward the initiation frequencies set \{$\omega_{max}>\omega_{m>}>\omega_{m<}>\omega_{min}$\}. The other two columns, in the middle, introduce the nearest neighbour frequency differences $\tilde{\Delta}_{max}>0$ and $\tilde{\Delta}_{m}<0$, respectively. The column to the right indicates the relocated phases of the oscillators.

Consequently, we have two choices, as they appear in the legends of the plots of Figure 2 (a and b) and in table II: the first is distinct when $\tilde{\omega}_{max}=\omega_{max}$, $\tilde{\omega}_{min}=\omega_{m<}$, $\tilde{\omega}_{m>}=\omega_{m>}$, and $\tilde{\omega}_{m<}=\omega_{min}$ (see the main plots of Figure 2a and 2b as well as the first sub-row, within any row in table II, for any configuration). The second is defined as $\tilde{\omega}_{max}=\omega_{m>}$, $\tilde{\omega}_{min}=\omega_{min}$, $\tilde{\omega}_{m>}=\omega_{max}$, and $\tilde{\omega}_{m<}=\omega_{m<}$ (see the upper inset of Figure 2a and the upper inset of Figure 2b in addition to the second sub-row, inside any row in table II, for any arrangement). Any of the two choices is fulfilled when the pattern 5 appears (the lower inset of Figure 2b). Accordingly, as shown in table II, we have two selections for each configuration, either $|\Delta_i^{nn}|_{max} \equiv \tilde{\Delta}_{max}=(\tilde{\omega}_{max}-\tilde{\omega}_{min}) \equiv (\omega_{max}-\omega_{m<})>0$ in addition to $(\Delta_i^{nn})_{m} \equiv \tilde{\Delta}_{m}=(\tilde{\omega}_{m<}-\tilde{\omega}_{m>}) \equiv (\omega_{min}-\omega_{m>})<0$ or $|\Delta_i^{nn}|_{max} \equiv \tilde{\Delta}_{max}=(\tilde{\omega}_{max}-\tilde{\omega}_{min}) \equiv (\omega_{m>}-\omega_{min})>0$ along with $\Delta_i^{nn}|_{m} \equiv \tilde{\Delta}_{m}=(\tilde{\omega}_{m<}-\tilde{\omega}_{m>}) \equiv (\omega_{m<}-\omega_{max})<0$. These two choices match the phases to be either $\tilde{\theta}_{max}=\theta_{max}$, $\tilde{\theta}_{m>}=\theta_{m>}$, $\tilde{\theta}_{min}=\theta_{m<}$ additional to $\tilde{\theta}_{m<}=\theta_{min}$ or $\tilde{\theta}_{max}=\theta_{m>}$, $\tilde{\theta}_{m>}=\theta_{max}$, $\tilde{\theta}_{min}=\theta_{min}$ as well as $\tilde{\theta}_{m<}=\theta_{m<}$, respectively. In this class II, we always notice the previously mentioned two selections that agree with the phase lock condition, at $K_c$, is either $\phi_{max}=(\theta_{max}-\theta_{m<})=\pi/2$ or $\phi_{max}=(\theta_{m>}-\theta_{min})=\pi/2$, in sequence. The phase lock condition match up whichever the frequency difference $\tilde{\Delta}_{max}=(\omega_{max}-\omega_{m<})>0$ or $\tilde{\Delta}_{max}=(\omega_{m>}-\omega_{min})>0$, respectively.

A comparison between tables I and II shows that the left columns in both tables include different configurations in each class. In addition, class II, contains more details in comparison to class I. These details appear because the arrangements of the initial frequencies over the set \{$\omega_{max}>\omega_{m>}>\omega_{m<}>\omega_{min}$\} do not fulfill the requirements that all the ordered oscillators, in each assembly, are nearest neighbours. The legends of all plots in Figure 2(a and b) clearly indicate these details. Therefore, in order to held the oscillators as nearest neighbours and hence the definitions of the frequency differences according to equations (1) and (2), we must redefine frequencies as they appear in the $1^{st}$ and $3^{rd}$ columns (from left) of table II. Also, in class II, the phases are defined in a different manner (the farthest right column in table II). Subsequently, in each column in table II, for each configuration, there are two sub-rows. This is because, in class II, we must include the two adoptions of the initial frequencies. Moreover, we have to embrace the two alternatives, for each configuration, of the nearest neighbour frequency difference $\tilde{\Delta}_{max}>0$ in addition to the nearest neighbour frequency difference $\tilde{\Delta}_{m}<0$. Also we present, for each arrangement, the two assortments of the phases as shown in the extreme right column.

\subsection{Class III}
\begin{figure}
\includegraphics[scale=0.5]{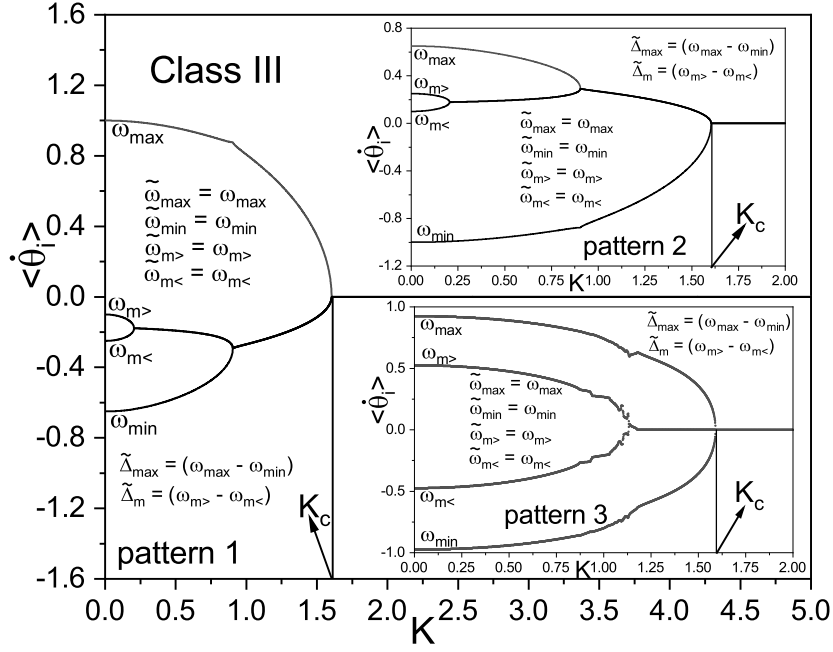}
\caption{\label{fig:three} Some exemplars of the synchronization diagrams of the non-identical four oscillators, in support of class III, as soon as the coupling constant increases for specific arrangements of \{$\omega_{max}>\omega_{m>}>\omega_{m<}>\omega_{min}$\}. Sceneries to realize dissimilar patterns are shown in the legend of each illustration. The major drawing refers to the pattern 1. The higher insertion graph indicates the pattern 2. The lower draw diagram implies the pattern 3. }
\end{figure}

Class III appears whenever the twosomes oscillators of $(\omega_{max} $\&$ \omega_{min})$ are nearest neighbour and the couples oscillators of $(\omega_{m>} $\&$ \omega_{m<})$ are nearest neighbour. Neither the combined oscillators of $(\omega_{max} $\&$ \omega_{m>})$ nor the other pair of $(\omega_{m<} $\&$ \omega_{min})$ are nearest neighbour oscillators. The principal illustration of Figure 3 (pattern 1), describes the synchronization tree, once the oscillator of $\omega_{max}$ meets the cluster of three oscillators of $\omega_{m>}$, $\omega_{m<}$ and $\omega_{min}$ at $K_c$. The upper insertion pattern of Figure 3 (pattern 2), put on view the synchronization feature as the cluster of three oscillators of $\omega_{max}$, $\omega_{m>}$ and $\omega_{m<}$ links to the oscillator of $\omega_{min}$ at the critical coupling value. The lower inset diagram of Figure 3 (pattern 3), displays the synchronization behaviour when the assembly of two oscillators of $\omega_{max}$ and $\omega_{min}$ joins the group of two oscillators of $\omega_{m>}$ and $\omega_{m<}$ at $K_c$. In class III, neither we obtain patterns similar to patterns 3, 4 and 6 of class I (see Figure 1(a and b)) nor we observe displays similar to patterns 3, 4 and 5 of class II (see Figure 2b). Also, within class III, we obtain a unique pattern 3 which is different from the pattern 5 in class I.

Furthermore, the pattern 3 in class III is not like to the patterns 5 in class I because the oscillators of $\omega_{max}$ and $\omega_{m>}$ as well as the oscillators of $\omega_{m<}$ and $\omega_{min}$ are not nearest neighbours in class III.
%In addition, within class III, the oscillators of $\omega_{m>}$ and $\omega_{min}$ω_(m>) and ω_min are not nearest neighbours.
Therefore, class III is obtained when the frequencies are $\tilde{\omega}_{max}=\omega_{max}$, $\tilde{\omega}_{min}=\omega_{min}$, $\tilde{\omega}_{m>}=\omega_{m>}$ and $\tilde{\omega}_{m<}=\omega_{m<}$. Moreover, the phases are regarded as $\tilde{\theta}_{max}=\theta_{max}$, $\tilde{\theta}_{min}=\theta_{min}$, $\tilde{\theta}_{m>}=\theta_{m>}$ and $\tilde{\theta}_{m<}=\theta_{m<}$.
Thus, we characterize class III by $|\tilde{\Delta}_i^{nn}|_{max} \equiv \tilde{\Delta}_{max}=(\omega_{max}-\omega_{min})>0$ and $(\tilde{\Delta}_i^{nn})_{m} \equiv \tilde{\Delta}_{m}=(\omega_{m>}-\omega_{m<})>0$.
The phase lock condition, for class III, $\phi_{max}=(\theta_{max}-\theta_{min})=\pi/2$ at $K_c$ corresponds to $\tilde{\Delta}_{max}>0$.  The different eight classifications of oscillators according to class III are itemized in table III. As stated in table III, the left column presents the different eight arrangements to assign the initial frequencies on the set \{$\omega_{max}>\omega_{m>}>\omega_{m<}>\omega_{min}$\}. The other two columns to the right explain the definitions of the nearest neighbour quantities $\tilde{\Delta}_{max}>0$ and $\tilde{\Delta}_{m}>0$.
\begin{table}
\caption{The eight compositions of class III. The definitions of both $\tilde{\Delta}_{max}>0$ and $\tilde{\Delta}_{m}>0$ are particularly set for each specific configuration.}
%{tabular environment}%
\label{table-three}
\begin{tabular}{|c|c|c|}
\hline
 configuration & $\tilde{\Delta}_{max}>0$ & $\tilde{\Delta}_{m}>0$  \\
$(\omega_{\max},\omega_{m>},\omega_{m<},\omega_{min})$  & $\tilde{\Delta}_{max}=(\tilde{\omega}_{max}-\tilde{\omega}_{min})$ &  $\tilde{\Delta}_{m}=(\tilde{\omega}_{m>}-\tilde{\omega}_{m<})$  \\
\hline
 $(\omega_1,\omega_3,\omega_4,\omega_2)$  & $\tilde{\Delta}_{max}=(\omega_1-\omega_2)$  & $\tilde{\Delta}_{m}=(\omega_3-\omega_4)$  \\
\hline
 $(\omega_1,\omega_3,\omega_2,\omega_4)$  & $\tilde{\Delta}_{max}=(\omega_1-\omega_4)$   & $\tilde{\Delta}_{m}=(\omega_3-\omega_2)$ \\
\hline
 $(\omega_2,\omega_4,\omega_1,\omega_3)$  & $\tilde{\Delta}_{max}=(\omega_2-\omega_3)$  & $\tilde{\Delta}_{m}=(\omega_4-\omega_1)$  \\
\hline
 $(\omega_2,\omega_4,\omega_3,\omega_1)$  & $\tilde{\Delta}_{max}=(\omega_2-\omega_1)$  & $\tilde{\Delta}_{m}=(\omega_4-\omega_3)$  \\
\hline
 $(\omega_3,\omega_1,\omega_4,\omega_2)$  & $\tilde{\Delta}_{max}=(\omega_3-\omega_2)$  & $\tilde{\Delta}_{m}=(\omega_1-\omega_4)$    \\
\hline
 $(\omega_3,\omega_1,\omega_2,\omega_4)$  & $\tilde{\Delta}_{max}=(\omega_3-\omega_4)$  & $\tilde{\Delta}_{m}=(\omega_1-\omega_2)$ \\
\hline
 $(\omega_4,\omega_2,\omega_1,\omega_3)$  & $\tilde{\Delta}_{max}=(\omega_4-\omega_3)$  & $\tilde{\Delta}_{m}=(\omega_2-\omega_1)$ \\
\hline
 $(\omega_4,\omega_2,\omega_3,\omega_1)$  & $\tilde{\Delta}_{max}=(\omega_4-\omega_1)$  & $\tilde{\Delta}_{m}=(\omega_2-\omega_3)$ \\
 \hline
\end{tabular}%
\end{table}

It should be noted that there are three major differences between class I and class III. The first exists directly from the different arrangements of the initial frequencies which appear within the left columns in both tables I and III. Moreover, the configurations in class III include some non-nearest neighbour oscillators as we order the initial frequencies over the set \{$\omega_{max}>\omega_{m>}>\omega_{m<}>\omega_{min}$\}. The second become visible because of the quantities $(\omega_{max}-\omega_{m>})$ and $(\omega_{m>}-\omega_{min})$ are not nearest neighbours in class III in comparison to the similar quantities in class I. We neglect the use of both quantities $(\omega_{max}-\omega_{m>})$ and $(\omega_{m>}-\omega_{min})$ in both classes I and III (as will be clear later). The third appears in the definition of the frequency difference $\tilde{\Delta}_{m}=(\omega_{max}-\omega_{min})>0$, in order to distinguish class III when we treat later equations (1) to arrive to an analytic solution in regard to this class III.
As we expect, the equation for calculating $K_c$ will be given in terms of $\tilde{\Delta}_{max}$ and $\tilde{\Delta}_{m}$. Therefore, $\tilde{\Delta}_{m}$ must have a different sign when it is operated in class III than that is used for class I. This change is necessary so as to distinguish the exact values of $K_c$ for class III from the values of the critical coupling of class I. Also, class III sets apart from class II as the comparison between tables II and III shows different configurations in each class (II and III). Additionally, the definitions of $\tilde{\Delta}_{max}$ and $\tilde{\Delta}_{m}$ are different for class III in comparison to the corresponding quantities in class II. Also, in class III, there is no need to redefine the phases as they are outlined in the extreme right column of table II of class II.

In all diagrams of Figures 1, 2 and 3, and their insets, we refer to $K_c$ in the graphs, where the oscillators start to synchronize to each other having equal $\dot\theta_i$, for $i = 1, 2, 3, 4$, as time independent quantities. The four quantities $\dot\theta_i$ own numerical values $< 10^{-5}$ and they remain at these values for an extremely long time. Also, we determine numerically $K_c$ when the phase differences $\phi_i$, for $i = 1, 2, 3, 4$, become time independent and they possess numerical values $< 10^{-5}$ as well as they persist at these values for an enormously long time. At $K_c$, system (3) undergoes a saddle-node bifurcation and the phase slip features of $\dot\theta_i$ and $\dot\theta_i$, for $i = 1, 2, 3, 4$, disappear.

\section{Exact Solution at The Instant of Synchronization}
As stated in the previous section, we find the following: the consecutive pairs of oscillators in each configuration of class I appear nearest neighbours. Consequently, the nearest neighbours quantities $\tilde{\Delta}_{max}=(\omega_{max}-\omega_{min})$ and $\tilde{\Delta}_{m}=(\omega_{m<}-\omega_{m>})$ are obtained. In class II, the two oscillators of $\omega_{max}$ and $\omega_{min}$ seem non-nearest neighbours as well as the other two oscillators of $\omega_{m>}$ and $\omega_{m<}$ look non-nearest neighbours. Therefore, in class II, we need to redefine the frequencies to allocate pairs of nearest neighbours as $(\tilde{\omega}_{max} \& \tilde{\omega}_{m>})$, $(\tilde{\omega}_{m>} \& \tilde{\omega}_{m<})$, $(\tilde{\omega}_{m<} \& \tilde{\omega}_{min})$ and $(\tilde{\omega}_{max} \& \tilde{\omega}_{min})$. Also, we redefine their corresponding phases to be $(\tilde{\theta}_{max} \& \tilde{\theta}_{m>})$, $(\tilde{\theta}_{m>} \& \tilde{\theta}_{m<})$, $(\tilde{\theta}_{m<} \& \tilde{\theta}_{min})$ and $(\tilde{\theta}_{max} \& \tilde{\theta}_{min})$. Henceforward, we define the nearest neighbour frequency differences $\tilde{\Delta}_{max}=(\tilde{\omega}_{max}-\tilde{\omega}_{min})$ and $\tilde{\Delta}_{m}=(\tilde{\omega}_{m<}-\tilde{\omega}_{m>})$. In class III, the oscillators of $(\omega_{max} \& \omega_{m>})$ and those of $(\omega_{m<} \& \omega_{min})$ are non-nearest neighbours. However, the pairs of $(\omega_{max} \& \omega_{min})$ and $(\omega_{m>} \& \omega_{m<})$ are nearest neighbours, which make the necessary frequency differences between the nearest neighbours as $\tilde{\Delta}_{max}=(\omega_{max}-\omega_{min})$ and $\tilde{\Delta}_{m}=(\omega_{m>}-\omega_{m<})$. Therefore, the major effective difference between class I and class III are coming due to the two nearest neighbour oscillators’ frequency difference $\tilde{\Delta}_{max}>0$. The reallocated quantities pave the way to obtain a unified solution to the three classes. Thus, for the three classes (see tables I, II and III), we write system (1) as
\begin{align}
\tilde{\omega}_{max}+\frac{K_c}{3}[\sin(\tilde{\theta}_{m>}-\tilde{\theta}_{max})
+\sin(\tilde{\theta}_{min}-\tilde{\theta}_{max})]=0, \nonumber \\
\tilde{\omega}_{m>}+\frac{K_c}{3}[\sin(\tilde{\theta}_{max}-\tilde{\theta}_{m>})
+\sin(\tilde{\theta}_{m<}-\tilde{\theta}_{m>})]=0, \nonumber \\
\tilde{\omega}_{m<}+\frac{K_c}{3}[\sin(\tilde{\theta}_{m>}-\tilde{\theta}_{m<})
+\sin(\tilde{\theta}_{min}-\tilde{\theta}_{m<})]=0, \nonumber \\
%\hspace{-20.2cm}\textrm{and} \nonumber \\
\tilde{\omega}_{min}+\frac{K_c}{3}[\sin(\tilde{\theta}_{m<}-\tilde{\theta}_{min})
+\sin(\tilde{\theta}_{max}-\tilde{\theta}_{min})]=0. \label{3}
\end{align}
Always for each configuration within each class, we use the first and the last two relations of system (3) that express the two oscillators possessing frequencies $\tilde{\omega}_{max}$ and $\tilde{\omega}_{min}$, to write the following equations
\begin{subequations}
\label{4}
\begin{eqnarray}
&&\sin(\tilde{\theta}_{m>}-\tilde{\theta}_{max})=1-\frac{3 \tilde{\omega}_{max}}{K_c}, \label{4a} \\
%and \nonumber \\
&&\sin (\tilde{\theta}_{min}-\tilde{\theta}_{m<})=1+\frac{3 \tilde{\omega}_{min}}{K_c}. \label{4b}
\end{eqnarray}
\end{subequations}
Moreover, we utilize the equations of system (3), at the synchronization point, to arrive at
\begin{subequations}
\label{5}
\begin{eqnarray}
&&\frac{3 \tilde{\Delta}_{max}}{K_c}+\sin (\tilde{\theta}_{m>}-\tilde{\theta}_{max})+\sin (\tilde{\theta}_{min}-\tilde{\theta}_{m<})=2, \label{5a} \\
%and \nonumber \\
&&\frac{3 \tilde{\Delta}_m}{K_c}+\sin (\tilde{\theta}_{m>}-\tilde{\theta}_{max})+\sin (\tilde{\theta}_{min}-\tilde{\theta}_{m<})=2\sin (\tilde{\theta}_{m>}-\tilde{\theta}_{m<}), \label{5b}
\end{eqnarray}
\end{subequations}
where, $\tilde{\omega}_{max}$, $\tilde{\omega}_{m>}$, $\tilde{\omega}_{m<}$, $\tilde{\omega}_{min}$, $\tilde{\Delta}_{max}$ and $\tilde{\Delta}_{m}$ are defined for any configuration within each class in tables I, II, and III. Also, the quantities $\tilde{\theta}_{max}$, $\tilde{\theta}_{m>}$, $\tilde{\theta}_{m<}$ and $\tilde{\theta}_{min}$ are demarcated for each class as specified in the previous section. We inscribe the expressions (4) and (5) using the phase lock condition $(\tilde{\theta}_{max}-\tilde{\theta}_{min})=\pi/2$ at $K_c$. Therefore, we apply the phase lock condition to obtain
\begin{align}
&\sin(\tilde{\theta}_{m>}-\tilde{\theta}_{m<})=-\cos (\tilde{\theta}_{m>}-\tilde{\theta}_{max}+\tilde{\theta}_{min}-\tilde{\theta}_{m<}), \nonumber \\
&\cos(\tilde{\theta}_{m>}-\tilde{\theta}_{max}+\tilde{\theta}_{min}-\tilde{\theta}_{m<})=
\cos(\tilde{\theta}_{m>}-\tilde{\theta}_{max})\cos(\tilde{\theta}_{min}-\tilde{\theta}_{m<}) \nonumber \\
&\hspace{+5cm}-\sin (\tilde{\theta}_{m>}-\tilde{\theta}_{max})\sin(\tilde{\theta}_{min}-\tilde{\theta}_{m<}), \nonumber \\
&\cos (\tilde{\theta}_{m>}-\tilde{\theta}_{max})=\sqrt{1-(1-\frac{3 \tilde{\omega}_{max}}{K_c})^2}, \nonumber \\
%and \nonumber \\
&\cos (\tilde{\theta}_{min}-\tilde{\theta}_{m<})=\sqrt{1-(1+\frac{3 \tilde{\omega}_{min}}{K_c})^2}. \nonumber
\end{align}
Accordingly, we draw on the directly above relations in addition to equations (4) and (5a) to arrive to the following expression
\begin{align}
&\sin (\tilde{\theta}_{m>}-\tilde{\theta}_{m<})= \nonumber \\
&\frac{1}{K_c^2}(-K_c^2+3 K_c \tilde{\Delta}_{max}+9 \tilde{\omega}_{max} \tilde{\omega}_{min}+3 \sqrt{\tilde{\omega}_{max} \tilde{\omega}_{min} (3 \tilde{\omega}_{max}-2 K_c) (2 K_c+3 \tilde{\omega}_{min})}). \label{6}
\end{align}
Also, we relate equations (5) and (6) together to yield the following mathematical form
\be
3 K_c (\tilde{\Delta}_{max}+\tilde{\Delta}_{m})+6 (\sqrt{\tilde{\omega}_{max } \tilde{\omega}_{min} (3 \tilde{\omega}_{max}-2 K_c) (3 \tilde{\omega}_{min}+2 K_c)}+3 \tilde{\omega}_{max} \tilde{\omega}_{min})=0.
\label{eq:seven}
\ee
Thus, we treat equation (7) to obtain the resulting equation, which is written as
\begin{align}
&-9 K_c(-K_c \tilde{\Delta}_{max}^2-K_c \tilde{\Delta}_m^2-2 K_c \tilde{\Delta}_{max} \tilde{\Delta}_{m}-16 K_c \tilde{\omega}_{max} \tilde{\omega}_{min}-12 \tilde{\Delta}_{max} \tilde{\omega}_{max} \tilde{\omega}_{min} \nonumber \\
&-12 \tilde{\Delta}_m \tilde{\omega}_{max} \tilde{\omega}_{min}
+24 \tilde{\omega}_{max}^2 \tilde{\omega}_{min}-24 \tilde{\omega}_{max} \tilde{\omega}_{min}^2)=0. \label{8}
\end{align}
When the following conditions are satisfied
\begin{align}
&K_c \neq 0, \nonumber \\
&\tilde{\Delta}_{max} \neq 0, \nonumber \\
&\tilde{\omega}_{max} \tilde{\omega}_{min} \neq 0, \nonumber \\
%and \nonumber \\
&(\tilde{\Delta}_{max}+\tilde{\Delta}_{m})^2+16 \tilde{\omega}_{max} \tilde{\omega}_{min} \neq 0, \label{9}
\end{align}
equation (8) gives exactly $K_c$, for classes I, II and III, that is written as
\begin{align}
&\frac{K_c}{\tilde{\Delta}_{max}}=1+\frac{\tilde{\Delta}_mH}{\tilde{\Delta}_{max}}, \nonumber \\
%where \nonumber \\
&H=-\frac{\tilde{\Delta}_{max} (\tilde{\Delta}_{max}+\tilde{\Delta}_{m})^2+4 \tilde{\omega}_{max} \tilde{\omega}_{min} (3 \tilde{\Delta}_{max}+\tilde{\Delta}_{m})}{\tilde{\Delta}_m((\tilde{\Delta}_{max}+\tilde{\Delta}_{m})^2
+16 \tilde{\omega}_{max} \tilde{\omega}_{min})}. \label{10}
\end{align}

\begin{figure}
\includegraphics[scale=0.35]{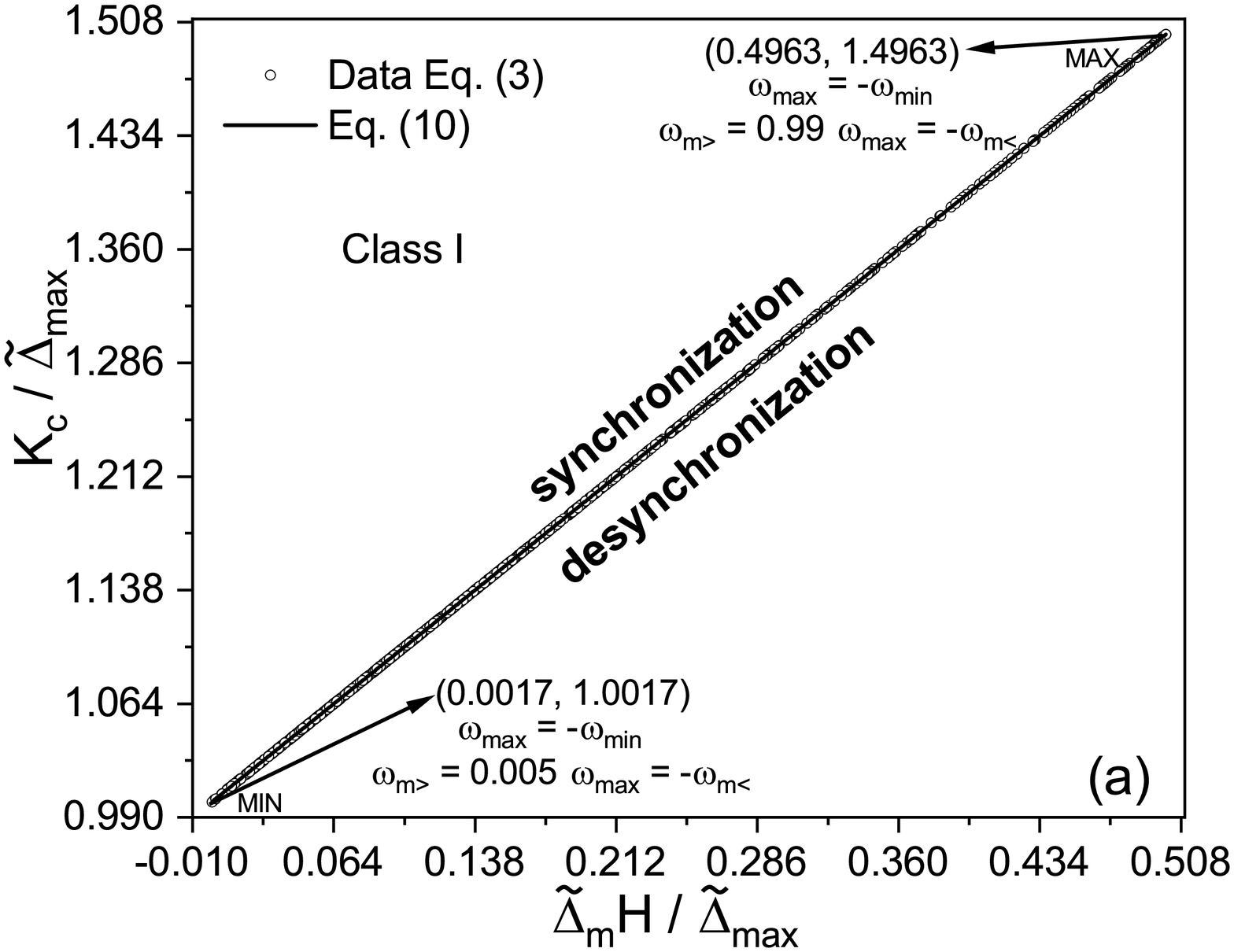}
\includegraphics[scale=0.35]{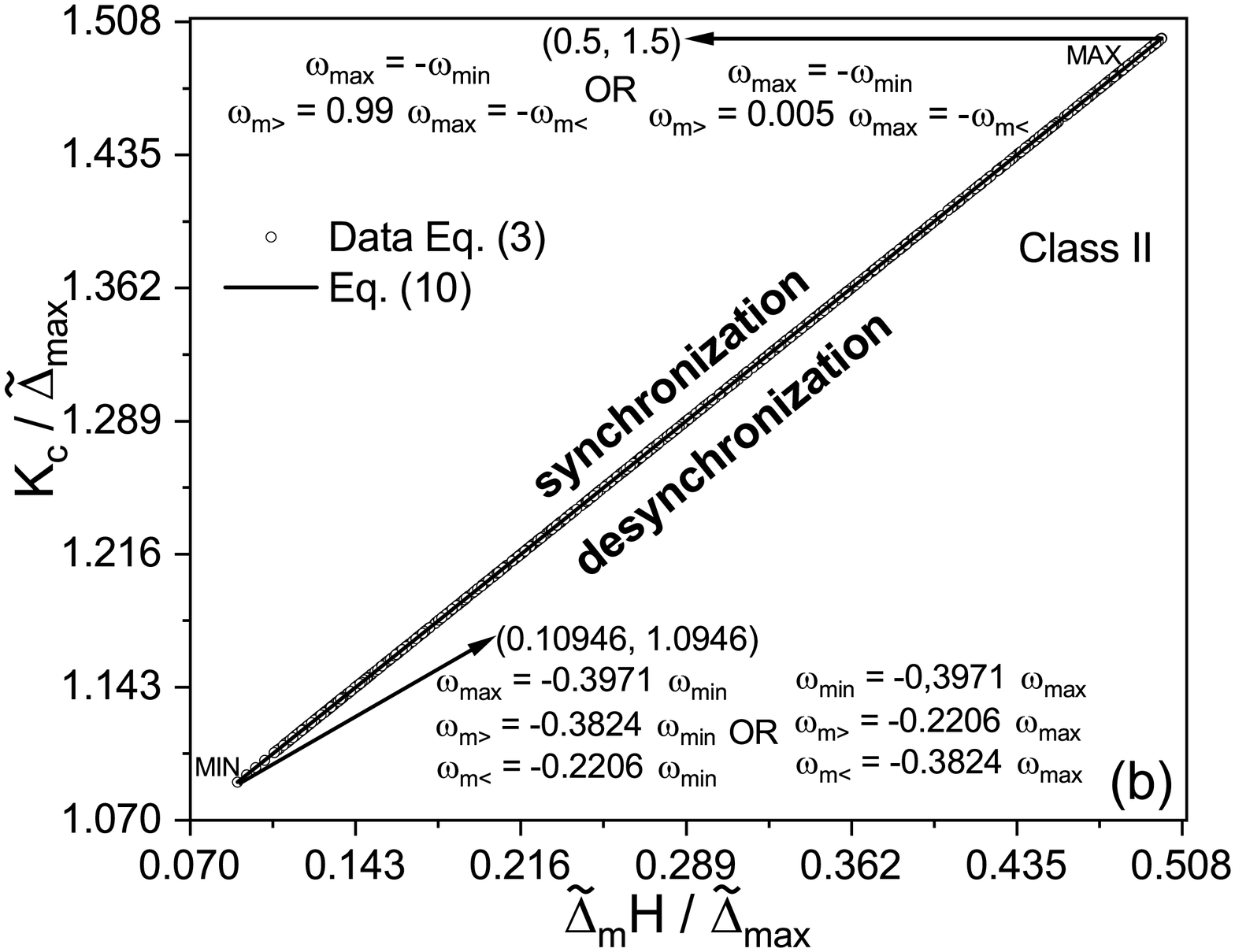}
\includegraphics[scale=0.35]{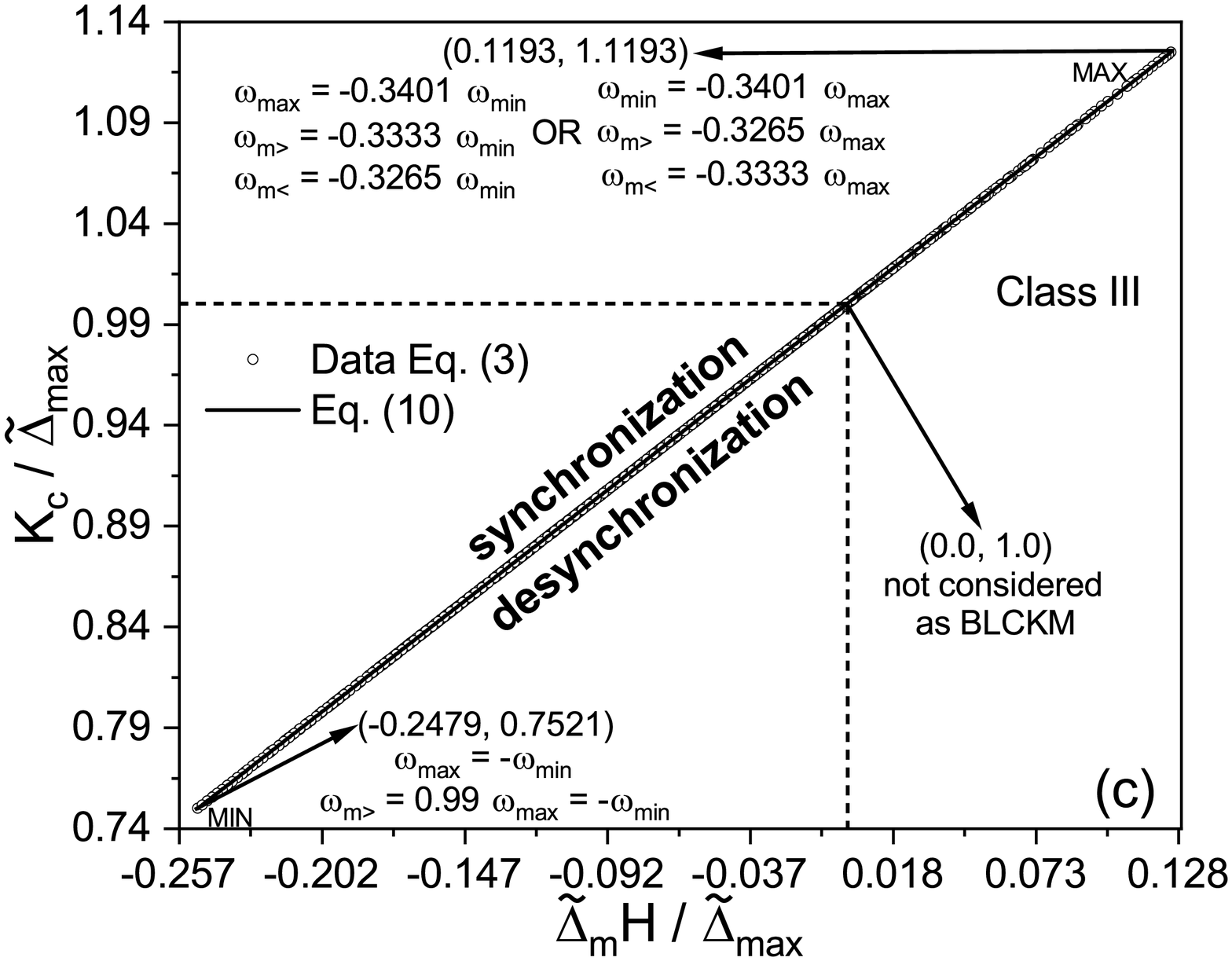}
\caption{\label{fig:four} The illustrations represent the transition from the desynchronization region to the synchronization zone for all classes. Data obtained numerically from system (3) are denoted by open circles while equation (10) is represented by the solid line in each drawing within each class. (a) The graph shows, for class I, the coincidence of the numerical data as circles with equation (10) as a solid line. (b) The plot depicts the justification of expression (10) in a connected line that fit well with the computed data as circles in the case of class II. (c) The diagram presents the matching between the calculated data as circles and formula (10) as a solid line in the occasion of class III. In each plot, the upper and lower arrows refer to the maximum and minimum limits for each class as well as the conditions of initial frequencies. Also, as shown in (c), the sign of the abscissa is changed from a negative value to a positive one at coordinates $(0.0, 1.0)$ as indicated by the middle arrow. }
\end{figure}

Thus, expression (10) determines the values of $K_c$ once we know the initial frequencies \{$\tilde{\omega}_{max}, \tilde{\omega}_{m>},\tilde{\omega}_{m<},\tilde{\omega}_{min}$\} of the four oscillators for any configuration within any class. According to equation (10), the analytically calculated values of $K_c$ depend on the quantities $\tilde{\Delta}_{max}$, $\tilde{\Delta}_{m}$, $\tilde{\omega}_{max}$ and $\tilde{\omega}_{min}$ as expected. In expression (10), the quantities $\tilde{\Delta}_{max}$ and $\tilde{\Delta}_{m}$ are considered as the frequency differences between nearest neighbour oscillators. Also, in formula (10), $\tilde{\omega}_{max}$ and $\tilde{\omega}_{min}$ are frequencies of nearest neighbour oscillators. Thus, equation (10) is serviceable, for each class (see tables I, II and III), once we know the initial frequencies $\tilde{\omega}_{max}$, $\tilde{\omega}_{min}$, $\tilde{\omega}_{m>}$ and $\tilde{\omega}_{m<}$ and define the frequency differences $\tilde{\Delta}_{max}$ and $\tilde{\Delta}_{m}$.

Figure 4 presents the confirmation of equation (10) to express $K_c$ for the three classes. Figures 4a, 4b and 4c show the plots of $K_c/\tilde{\Delta}_{max}$ versus $\tilde{\Delta}_mH/\tilde{\Delta}_{max}$, where $H$ is given in expression (10). Figure 4 shows clearly a complete agreement between $K_c$ (analytic) and $K_c$ (calculated numerically). The computed data, obtained by using equation (3), are shown as open white circles. These data match excellently the solid line (for each class) that represents equation (10). Figures 4a and 4b have positive abscissas in the cases of classes I and II but the definitions of the quantities $\tilde{\Delta}_{max}$ and $\tilde{\Delta}_{m}$ are different in each class. For each configuration in class II, $\tilde{\omega}_{max}$, $\tilde{\omega}_{min}$, $H$ and $K_c$ are possessing different values in contrast to the similar quantities in class I (the legends of Figure 4a and 4b present examples). In addition, the configurations \{$\tilde{\omega}_{max}, \tilde{\omega}_{m>},\tilde{\omega}_{m<},\tilde{\omega}_{min}$\} in both classes I and II are different. Figure 4c represents the validation of equation (10) for class III. Figure 4c appears different in comparison to figures 4a and 4b, because class III contains the quantity $\tilde{\Delta}_{m}>0$. When $\tilde{\Delta}_{m}>0$ is used in equation (10), we obtain the values of $K_c$ that matches class III. This is because $K_c$ in equation (10) depends on $\tilde{\Delta}_{max}>0$, $\tilde{\Delta}_{m}>0$, $\tilde{\omega}_{max}$ and $\tilde{\omega}_{min}$ that lead to have different abscissa and ordinate in the case of Figure 4c of class III. Moreover, in Figure 4c, the x-axis is ranged from a negative value to a positive one. This is for the reason that the quantity $H(\tilde{\Delta}_{max},\tilde{\Delta}_{m},\tilde{\omega}_{max},\tilde{\omega}_{min})$  changes its sign, depending on the values of the initial frequencies, while $\tilde{\Delta}_{m}$ is always positive. In Figure 4(a-c), for each class, the area under the solid line represents the region where the oscillators are in desynchronized states while the zone above the solid line signifies the synchronized states. As shown in the diagrams of Figure 4, any different coordinates denote different cases of $K_c$, $\tilde{\Delta}_{max}$, $\tilde{\Delta}_{m}$ and $H$ for each class.

\section{Further Elucidations Concerning The Critical Coupling}
According to Figure 4(a, b and c), the minimum and maximum limits in the three plots are different. These differences make an additional distinction of each class. Class I possess a minimum limit at the coordinates $(\tilde{\Delta}_mH/\tilde{\Delta}_{max}\approx0.0017,K_c/\tilde{\Delta}_{max}\approx1.0017)$ when the assigning of the initial frequencies are $\omega_{max}=-\omega_{min}$ and $\omega_{m>}=0.005\omega_{max}=-\omega_{m<}$. The maximum boundary, for class I, appears at the coordinates $(\tilde{\Delta}_mH/\tilde{\Delta}_{max}\approx0.4963,K_c/\tilde{\Delta}_{max}\approx1.4963)$ once the starting frequencies are allocated to be $\omega_{max}=-\omega_{min}$ and $\omega_{m>}=0.99\omega_{max}=-\omega_{m<}$. For class II, the minimum border is located at $(\tilde{\Delta}_mH/\tilde{\Delta}_{max}\approx0.10946,K_c/\tilde{\Delta}_{max}\approx1.0946)$ whenever we find either the case $\omega_{max}=-0.3971\omega_{min}$, $\omega_{m>}=-0.3824\omega_{min}$ as well as $\omega_{m<}=-0.2206\omega_{min}$ or the situation $\omega_{min}=-0.3971\omega_{max}$, $\omega_{m>}=-0.2206\omega_{max}$ along with $\omega_{m<}=-0.3824\omega_{max}$. It is not possible, upon taking into considerations the different configurations of class II, to have a lower limit than the previously observed limit. This is because the quantity
$\tilde{\Delta}_{m}$ does not own a less negative value close to zero or being zero. The upper bound, for class II, occurs at $(\tilde{\Delta}_mH/\tilde{\Delta}_{max}=0.5,K_c/\tilde{\Delta}_{max}=1.5)$ each time we have $\omega_{max}=-\omega_{min}$, $\omega_{m>}=0.99\omega_{max}$ in addition to $\omega_{m<}=-\omega_{m>}$ or we get a case of initial frequencies $\omega_{max}=-\omega_{min}$, $\omega_{m>}=0.005\omega_{max}$ and $\omega_{m<}=-\omega_{m>}$. Regarding class III, the lowest limit comes across at $(\tilde{\Delta}_mH/\tilde{\Delta}_{max}\approx-0.2479,K_c/\tilde{\Delta}_{max}\approx0.7521)$ as soon as the initial frequencies are allotted to be $\omega_{max}=-\omega_{min}$, $\omega_{m>}=0.99\omega_{max}$ additional to $\omega_{m<}=-\omega_{m>}$. The highest limit, in class III, ensues at $(\tilde{\Delta}_mH/\tilde{\Delta}_{max}\approx0.1193,K_c/\tilde{\Delta}_{max}\approx1.1193)$ when the preliminary frequencies are selected to be whichever $\omega_{max}=-0.3401\omega_{min}$, $\omega_{m>}=-0.3333\omega_{min}$ as well as $\omega_{m<}=-0.3265\omega_{min}$ or $\omega_{min}=-0.3401\omega_{max}$, $\omega_{m>}=-0.3265\omega_{max}$ in addition to $\omega_{m<}=-0.3333\omega_{max}$. We notice that the point of changing signs at the abscissa in Figure 4c occurs at $(0.0,1.0)$, when $\omega_{max}=-\omega_{min}$ and a case of two equal initial frequencies are coincident $\omega_{m>}=\omega_{m<}=0$, but this case does not belong to the BLCKM used in this work. Cases of x-coordinate to the left of the inversion point occur when $(\tilde{\Delta}_mH/\tilde{\Delta}_{max}<0$ and $K_c/\tilde{\Delta}_{max}<1$ $(H < 0)$. The cases of x-coordinate to the right of the overturn point appear once $(\tilde{\Delta}_mH/\tilde{\Delta}_{max}>0$ and $K_c/\tilde{\Delta}_{max}>1$ $(H > 0)$. This is credited to the verity that the upturn point represents a transition situation between the pattern 3 (see Figure 3) as well as its similar patterns, and both the cases of patterns 1 in addition to 2 (see Figure 3) besides their alike. The minimum critical coupling constants for the three classes are $K_{c-min}^I\approx1.0017\tilde{\Delta}_{max}$, $K_{c-min}^{II}\approx1.0946\tilde{\Delta}_{max}$ and $K_{c-min}^{III}\approx0.7521\tilde{\Delta}_{max}$. The maximum critical coupling constants for classes I, II and III are
$K_{c-max}^I\approx1.4963\tilde{\Delta}_{max}$, $K_{c-max}^{II}=1.5\tilde{\Delta}_{max}$ and $K_{c-max}^{III}\approx1.1193\tilde{\Delta}_{max}$. As we notice, class III has the lowest critical coupling constant while class II holds the largest value of the critical coupling constant. Also, the values of $K_{c-max}^I$ and $K_{c-max}^{II}$ are comparable as well as they occur at similar patterns of $\omega_{max}=-\omega_{min}$ and $\omega_{m>}=0.99\omega_{max}=-\omega_{m<}$. However, the value of $K_{c-max}^I$ can not increase more to be the same as $K_{c-max}^{II}$ because of, in case of non-identical BLCKM, the restriction that the frequency differences between any two or more oscillators must not have values less than $0.01$.

We also notice that $K_{c-max}^{I}$ and $K_{c-min}^{I}$ occur at conditions (as indicated in the legends of Figure 4a) when the patterns of synchronization trees show two oscillators of initial frequencies (relative to $\omega_o=0$) remaining up and two oscillators of initial frequencies staying down (comparative to $\omega_o=0$). However, $K_{c-min}^{I}$ takes place (see Figure 4a) at a pattern possessing $\omega_{m>}=-\omega_{m<}$ (they are close to $\omega_o=0$) while $K_{c-max}^{I}$ comes about for a pattern owing $\omega_{m>}=-\omega_{m<}$ (they are far from each other). For class II, $K_{c-max}^{II}$ (see plots of Figure 4(a and b)) comes to pass at conditions of a pattern that shows both $K_{c-max}^{I}$ and $K_{c-min}^{I}$ of class I. In class III, $K_{c-min}^{III}$ happens (see Figure 4c) at conditions that lead to have $K_{c-max}^{I}$. The value of $K_{c-min}^{II}$ ensues (see Figure 4b) as the patterns of synchronization trees show either one oscillator of initial frequency $\omega_{max}$ placing up and three oscillators of initial frequencies $\omega_{m>}+\omega_{m<}+\omega_{min}=-\omega_{max}$ holding down (relative to $\omega_o=0$) or three oscillators of initial frequencies $\omega_{max}+\omega_{m>}+\omega_{m<}=-\omega_{min}$ locating up and one oscillator of initial frequency $\omega_{min}$ lasting down (relative to $\omega_o=0$). We find $K_{c-max}^{III}$ (see Figure 4c) when a similar pattern that leads to $K_{c-min}^{II}$ of class II but the initial frequencies are different.

For any configuration in a certain class, we are familiar with the relationships between the frequencies $\tilde{\omega}_{max}$, $\tilde{\omega}_{min}$, $\tilde{\omega}_{m>}$ as well as $\tilde{\omega}_{m<}$ and their corresponding initial frequencies $\omega_{max}$, $\omega_{min}$, $\omega_{m>}$ in addition to $\omega_{m<}$. Also, for every class, we know how to express the frequency differences $\tilde{\Delta}_{max}$ along with $\tilde{\Delta}_{m}$ in terms of the initial frequencies $\omega_{max}$, $\omega_{min}$, $\omega_{m>}$ and $\omega_{m<}$. Accordingly, for class I, we define $\tilde{\omega}_{max}=\omega_{max}$, $\tilde{\omega}_{min}=\omega_{min}$, $\tilde{\omega}_{m>}=\omega_{m>}$, $\tilde{\omega}_{m<}=\omega_{m<}$, $\tilde{\Delta}_{max}=(\omega_{max}-\omega_{min})$ and $\tilde{\Delta}_{m}=(\omega_{m<}-\omega_{m>})$. Consequently, we write the critical coupling (equation (10)) in terms of the initial frequencies of each configuration of class I as
\begin{align}
&K_c^I=(\omega_{max}-\omega_{min})+(\omega_{m<}-\omega_{m>})H^I, \nonumber \\
&H^I=-\frac{A_1+B_1}{C_1}, \nonumber \\
&A_1=(\omega_{max}-\omega_{min})((\omega_{max}-\omega_{min})+(\omega_{m<}-\omega_{m>}))^2, \nonumber \\
&B_1=4((\omega_{max}-\omega_{min})+3(\omega_{m<}-\omega_{m>}))\omega_{max}\omega_{min}, \nonumber \\
%and \nonumber \\
&C_1=(\omega_{m<}-\omega_{m>})(((\omega_{max}-\omega_{min})+(\omega_{m<}-\omega_{m>}))^2+16 \omega_{max}
\omega_{min}). \label{11}
\end{align}
Within class III, For any configuration, we find $\tilde{\Delta}_{max}=(\omega_{max}-\omega_{min})$ and $\tilde{\Delta}_{m}=(\omega_{m>}-\omega_{m<})$. Also, the frequencies become $\tilde{\omega}_{max}=\omega_{max}$, $\tilde{\omega}_{min}=\omega_{min}$, $\tilde{\omega}_{m>}=\omega_{m>}$ and $\tilde{\omega}_{m<}=\omega_{m<}$. As a result, we express the critical coupling in terms of the initial frequencies for each configuration in class III as
\begin{align}
&K_c^{III}=((\omega_{max}-\omega_{min})+(\omega_{m>}-\omega_{m<})H^{III}, \nonumber \\
&H^{III}=-\frac{A_3+B_3}{C_3}, \nonumber \\
&A_3=(\omega_{max}-\omega_{min})((\omega_{max}-\omega_{min})+(\omega_{m>}-\omega_{m<}))^2, \nonumber \\
&B_3=4((\omega_{max}-\omega_{min})+3(\omega_{m>}-\omega_{m<}))\omega_{max}\omega_{min}, \nonumber \\
%and \nonumber \\
&C_3=(\omega_{m>}-\omega_{m<})(((\omega_{max}-\omega_{min})+(\omega_{m>}-\omega_{m<}))^2+16 \omega_{max}
\omega_{min}). \label{12}
\end{align}
It is evident in equations (11) and (12) that the difference, in determining $K_c$, between classes I and III appear due to $(\omega_{m<}-\omega_{m>})$ for class I and $(\omega_{m>}-\omega_{m<})$ for class III.

In every arrangements of class II, once the frequencies turn into $\tilde{\omega}_{max}=\omega_{max}$, $\tilde{\omega}_{min}=\omega_{m<}$, $\tilde{\omega}_{m>}=\omega_{m>}$ and $\tilde{\omega}_{m<}=\omega_{min}$. Additionally $\tilde{\Delta}_{max}=(\omega_{max}-\omega_{m<})$, $\tilde{\Delta}_{m}=(\omega_{min}-\omega_{m>})$ and $(\omega_{max}-\omega_{m<})>(\omega_{m>}-\omega_{min})$. Subsequently, we obtain the critical coupling in terms of the initial frequencies as
\begin{align}
&K_c^{II-a}=(\omega_{max}-\omega_{m>})+(\omega_{min}-\omega_{m>})H^{II-a}, \nonumber \\
&H^{II-a}=-\frac{A_{2-a}+B_{3-a}}{C_{3-a}}, \nonumber \\
&A_{2-a}=(\omega_{max}-\omega_{m>})((\omega_{max}-\omega_{m>})+(\omega_{min}-\omega_{m>}))^2, \nonumber \\
&B_{2-a}=4((\omega_{max}-\omega_{m>})+3(\omega_{min}-\omega_{m>}))\omega_{max}\omega_{m<}, \nonumber \\
%and \nonumber \\
&C_{2-a}=(\omega_{min}-\omega_{m>})(((\omega_{max}-\omega_{m>})+(\omega_{min}-\omega_{m>}))^2
+16\omega_{max}\omega_{m<}). \label{13}
\end{align}
Within class II, as soon as the frequencies held $\tilde{\omega}_{max}=\omega_{m>}$, $\tilde{\omega}_{min}=\omega_{min}$, $\tilde{\omega}_{m>}=\omega_{max}$ and $\tilde{\omega}_{m<}=\omega_{m<}$. Also, when we find $\tilde{\Delta}_{max}=(\omega_{m>}-\omega_{min})$, $\tilde{\Delta}_{m}=(\omega_{m<}-\omega_{max})$ and $(\omega_{m>}-\omega_{min})>(\omega_{m<}-\omega_{max})$. In this case, the critical coupling is written as
\begin{align}
&K_c^{II-b}=(\omega_{max}-\omega _{min})+(\omega_{m<}-\omega_{max})H^{II-b}, \nonumber \\
&H^{II-b}=-\frac{A_{2-b}+B_{3-b}}{C_{3-b}}, \nonumber \\
&A_{2-b}=(\omega_{m>}-\omega_{min})((\omega_{m>}-\omega_{min})+(\omega_{m<}-\omega_{max }))^2, \nonumber \\
&B_{2-b}=4((\omega_{m>}-\omega_{min})+3(\omega_{m<}-\omega_{max}))\omega_{m>}\omega_{min}, \nonumber \\
%and \nonumber \\
&C_{2-b}=(\omega_{m<}-\omega_{max})(((\omega_{m>}-\omega_{min})+(\omega_{m<}-\omega_{max}))^2
+16\omega_{m>}\omega_{min}). \label{14}
\end{align}

It is obvious from equations (11), (12), (13) and (14) that the value of the critical couplings depend on the initial nearest neighbour frequencies.
The minimum critical coupling constants for the three classes are $K_{c-min}^I\approx1.0017(\omega_{max}-\omega_{min})$, $K_{c-min}^{III}\approx0.7521(\omega_{max}-\omega_{min})$ and $K_{c-min}^{II-a}\approx1.0946(\omega_{max}-\omega_{m<})$ or $K_{c-min}^{II-b}\approx1.0946(\omega_{m>}-\omega_{min})$. The maximum critical coupling constants for classes I, III and II are $K_{c-max}^I\approx1.4963(\omega_{max}-\omega_{min})$, $K_{c-max}^{III}\approx1.1193(\omega_{max}-\omega_{min})$ and $K_{c-max}^{II-a}=1.5(\omega_{max}-\omega_{min})$ or $K_{c-max}^{II-b}=1.5(\omega_{m>}-\omega_{min})$. The minimum and maximum critical couplings $K_{c-min}^{II-a}$ and $K_{c-max}^{II-a}$ appear depending on the condition $(\omega_{max}-\omega_{m<})>(\omega_{m>}-\omega_{min})$. Also, the minimum and maximum critical couplings $K_{c-min}^{II-b}$ and $K_{c-max}^{II-b}$ occur relying on $(\omega_{m>}-\omega_{min})>(\omega_{max}-\omega_{m<})$.

\section{Phase Differences at The Critical Coupling}
System (3), the phase lock condition (4) and equation (10) allow us to calculate the phase differences at the moment the four local coupled phase oscillators synchronize having a common frequency. Therefore, for classes I, II and III, we write the phase differences as
\begin{align}
&\sin((\tilde{\theta}_{max}-\tilde{\theta}_{min}))=1, \nonumber \\
&\sin(\tilde{\theta}_{m>}-\tilde{\theta}_{max})=1-\frac{3 \tilde{\omega}_{max}}{H\tilde{\Delta}_{m}+\tilde{\Delta}_{max}}, \nonumber \\
&\sin(\tilde{\theta}_{min}-\tilde{\theta}_{m<})=1+\frac{3 \tilde{\omega}_{min}}{H\tilde{\Delta}_{m}+\tilde{\Delta}_{max}}, \nonumber \\
%and \nonumber \\
&\sin(\tilde{\theta}_{m<}-\tilde{\theta}_{m>})=\frac{H\tilde{\Delta}_{m}+\tilde{\Delta}_{max}-3(\tilde{\omega}_{m>}
+\tilde{\omega}_{max})}{H\tilde{\Delta}_{m}+\tilde{\Delta}_{max}} \nonumber \\
&\equiv \frac{H\tilde{\Delta}_{m}+\tilde{\Delta}_{max}
+3(\tilde{\omega}_{m<}+\tilde{\omega}_{min})}{H\tilde{\Delta}_{m}+\tilde{\Delta}_{max}}, \label{15}
\end{align}
where, $H$ is specified in equation (10). Relations (15) determine the phase differences $(\tilde{\theta}_{max}-\tilde{\theta}_{min})$, $(\tilde{\theta}_{m>}-\tilde{\theta}_{max})$, $(\tilde{\theta}_{min}-\tilde{\theta}_{m<})$ and $(\tilde{\theta}_{m<}-\tilde{\theta}_{m>})$ for any arrangement in each class. In particular, expression (15) is valid, within every class, when the conditions defining $\tilde{\Delta}_{max}$, $\tilde{\Delta}_{m}$, $\tilde{\omega}_{max}$, $\tilde{\omega}_{m>}$, $\tilde{\omega}_{m<}$, $\tilde{\omega}_{min}$, $\tilde{\theta}_{max}$, $\tilde{\theta}_{m>}$, $\tilde{\theta}_{m<}$ and $\tilde{\theta}_{min}$ are satisfied as specified in tables I, II and III.

We know, within each class, the connections between the frequencies $\tilde{\omega}_{max}$, $\tilde{\omega}_{m>}$, $\tilde{\omega}_{m<}$, added to $\tilde{\omega}_{min}$ and their consistent preliminary frequencies $\omega_{max}$, $\omega_{m>}$, $\omega_{m<}$ as well as $\omega_{min}$. Also, we are aware of, for any class, how to determine the frequency differences $\tilde{\Delta}_{max}$ along with $\tilde{\Delta}_{m}$ in terms of the initial frequencies $\omega_{max}$, $\omega_{m>}$, $\omega_{m<}$ and $\omega_{min}$. Also, we are acquainted with the relations (as defined for each class) between the phases $\tilde{\theta}_{max}$, $\tilde{\theta}_{m>}$, $\tilde{\theta}_{m<}$ in addition to $\tilde{\theta}_{min}$ and their corresponding phases $\theta_{max}$, $\theta_{m>}$, $\theta_{m<}$ together with $\theta_{min}$. Accordingly, we determine the phase differences $(\theta_{i+1}-\theta_i)$, for $i = 1, 2, 3, 4$.

Therefore, we write (3) (see tables I and III along with the conditions valid for any arrangements \{$\omega_{max}>\omega_{m>}>\omega_{m<}>\omega_{min}$\}), for classes I and III, as
\begin{align}
&\omega_{max}+\frac{K_c}{3}[\sin(\theta_{m>}-\theta_{max})+\sin(\theta_{min}-\theta_{max})]=0, \nonumber \\
&\omega_{m>}+\frac{K_c}{3}[\sin(\theta_{max}-\theta_{m>})+\sin(\theta_{m<}-\theta_{m>})]=0, \nonumber \\
&\omega_{m<}+\frac{K_c}{3}[\sin(\theta_{m>}-\theta_{m<})+\sin(\theta_{min }-\theta _{m<})]=0, \nonumber \\
%and \nonumber \\
&\omega_{min}+\frac{K_c}{3}[\sin(\theta_{m<}-\theta_{min})+\sin(\theta_{max}-\theta_{min})]=0, \label{16}
\end{align}
where, in (16), $K_c \equiv K_c^I$ for class I and $K_c \equiv K_c^{III}$ for class III. Thus (15), for class I as the phase terms appear in (16), can be written as
\begin{align}
&\sin(\theta_{max}-\theta_{min})=1, \nonumber \\
&\sin(\theta_{m>}-\theta_{max})=1-\frac{3 \omega_{max}}{H^I(\omega_{m<}-\omega_{m>})+\omega_{max}-\omega_{min}}, \nonumber \\
&\sin(\theta_{min}-\theta_{m<})=1+\frac{3 \omega_{min}}{H^I(\omega_{m<}-\omega_{m>})+\omega_{max}-\omega_{min}}, \nonumber \\
%and \nonumber \\
&\sin(\theta_{m<}-\theta_{m>})=\frac{H^I(\omega_{m<}-\omega_{m>})+\omega_{max}-\omega_{min}-3(\omega_{m>}
+\omega_{max})}{H^I(\omega_{m<}-\omega_{m>})+\omega_{max}-\omega_{min}} \nonumber \\
& \equiv \frac{H^I(\omega_{m<}-\omega_{m>})
+\omega_{max}-\omega_{min}+3(\omega_{m<}+\omega_{min})}{H^I(\omega_{m<}-\omega_{m>})+\omega_{max}-\omega_{min}}, \label{17}
\end{align}
where $H^I$ is given in equation (11).
The phase differences for class III, when we consider the phase terms in (16), are given as
\begin{align}
&\sin(\theta_{max}-\theta_{min})=1, \nonumber \\
&\sin(\theta_{m>}-\theta_{max})=1-\frac{3 \omega_{max}}{H^{III}(\omega_{m>}-\omega_{m<})+\omega_{max}-\omega_{min}}, \nonumber \\
&\sin(\theta_{min}-\theta_{m<})=1+\frac{3 \omega_{min}}{H^{III}(\omega_{m>}-\omega_{m<})+\omega_{max}-\omega_{min}}, \nonumber \\
%and \nonumber \\
&\sin(\theta_{m<}-\theta_{m>})=\frac{H^{III}(\omega_{m>}-\omega_{m<})+\omega_{max}-\omega_{min}-3(\omega_{m>}
+\omega_{max})}{H^{III}(\omega_{m>}-\omega_{m<})+\omega_{max}-\omega_{min}} \nonumber \\
& \equiv \frac{H^{III}(\omega_{m>}-\omega_{m<})
+\omega_{max}-\omega_{min}+3(\omega_{m>}+\omega_{min})}{H^{III}(\omega_{m>}-\omega_{m<})+\omega_{max}-\omega_{min}}, \label{18}
\end{align}
where $H^{III}$ is written in equation (12). In equations (17) and (18), we find $\Delta_m=(\omega_{m<}-\omega_{m>})$ for class I while $\Delta_m=(\omega_{m>}-\omega_{m<})$ for class III.

Regarding class II (see table II and necessities for any predeterminations \{$\omega_{max}>\omega_{m>}>\omega_{m<}>\omega_{min}$\}) there are two choices: the first comes out when the frequencies are $\tilde{\omega}_{max}=\omega_{max}$, $\tilde{\omega}_{min}=\omega_{m<}$, $\tilde{\omega}_{m>}=\omega_{m>}$ and $\tilde{\omega}_{m<}=\omega_{min}$. The corresponding phases are $\tilde{\theta}_{max}=\theta_{max}$, $\tilde{\theta}_{m>}=\theta_{m>}$, $\tilde{\theta}_{min}=\theta_{m<}$ along with $\tilde{\theta}_{m<}=\theta_{min}$. Thus, for the first choice in class II (refer to the first sub-row in each row of table II and prerequisites for any composition of the frequencies \{$\omega_{max}>\omega_{m>}>\omega_{m<}>\omega_{min}$\}, we inscribe (3) as
\begin{align}
&\omega_{max}+\frac{K_c^{II-a}}{3}[\sin(\theta_{m>}-\theta_{max})+\sin(\theta_{m<}-\theta_{max})]=0, \nonumber \\
&\omega_{m>}+\frac{K_c^{II-a}}{3}[\sin(\theta_{max}-\theta_{m>})+\sin(\theta_{min}-\theta_{m>})]=0, \nonumber \\
&\omega_{min}+\frac{K_c^{II-a}}{3}[\sin(\theta_{m>}-\theta_{min})+\sin(\theta_{m<}-\theta_{min})]=0, \nonumber \\
%and \nonumber \\
&\omega_{m<}+\frac{K_c^{II-a}}{3}[\sin(\theta_{min}-\theta_{m<})+\sin(\theta_{max}-\theta_{m<})]=0. \label{19}
\end{align}
The reorganization of (19) is necessary to obtain $\tilde{\Delta}_{max}=(\omega_{max}-\omega_{min})$ and $\tilde{\Delta}_{m}=(\omega_{min}-\omega_{m>})$ as differences between nearest neighbours oscillators. These lead, for the first choice (once the phase terms in (19) is taken into account), to write down (15) as
\begin{align}
&\sin(\theta_{max}-\theta_{m<})=1, \nonumber \\
&\sin(\theta_{m>}-\theta_{max})=1-\frac{3 \omega_{max}}{H^{II-a}(\omega_{min}-\omega_{m>})+(\omega_{max}-\omega_{m<})}, \nonumber \\
&\sin(\theta_{m<}-\theta_{min})=1+\frac{3\omega_{m<}}{H^{II-a}(\omega_{min}-\omega_{m>})+(\omega_{max}
-\omega_{m<})}, \nonumber \\
%and \nonumber \\
&\sin(\theta_{min}-\theta_{m>})=\frac{H^{II-a}(\omega_{min}-\omega_{m>})+(\omega_{max}
-\omega_{m<})-3(\omega_{m>}+\omega_{max})}{H^{II-a}(\omega_{min}-\omega_{m>})+(\omega_{max}
-\omega_{m<})} \nonumber \\
& \equiv \frac{H^{II-a}(\omega_{min}-\omega_{m>})+(\omega_{max}-\omega_{m<})
+3(\omega_{min}+\omega_{m<})}{H^{II-a}(\omega_{min}-\omega_{m>})+(\omega_{max}-\omega_{m<})}, \label{20}
\end{align}
where $H^{II-a}$ is given in equation (13). The second selection, for class II (see table II and requisites for any assemblages of \{$\omega_{max}>\omega_{m>}>\omega_{m<}>\omega_{min}$\}) arises whenever the frequencies are $\tilde{\omega}_{max}=\omega_{m>}$, $\tilde{\omega}_{min}=\omega_{min}$, $\tilde{\omega}_{m>}=\omega_{max}$ and $\tilde{\omega}_{m<}=\omega_{m<}$. The consistent phases are $\tilde{\theta}_{max}=\theta_{m>}$, $\tilde{\theta}_{m>}=\theta_{max}$, $\tilde{\theta}_{min}=\theta_{min}$ as well as $\tilde{\theta}_{m<}=\theta_{m<}$. Hence, refer to the second sub-row in each row of table II and criterions for any configuration of the frequencies \{$\omega_{max}>\omega_{m>}>\omega_{m<}>\omega_{min}$\}, we write out (3) as
\begin{align}
&\omega_{m>}+\frac{K_c^{II-b}}{3}[\sin(\theta_{max}-\theta_{m>})+\sin(\theta_{min}-\theta_{m>})]=0, \nonumber \\
&\omega_{max}+\frac{K_c^{II-b}}{3}[\sin(\theta_{m>}-\theta_{max})+\sin(\theta_{m<}-\theta_{max})]=0, \nonumber \\
&\omega_{m<}+\frac{K_c^{II-b}}{3}[\sin(\theta_{max}-\theta_{m<})+\sin(\theta_{min}-\theta_{m<})]=0, \nonumber \\
%and \nonumber \\
&\omega_{min}+\frac{K_c^{II-b}}{3}[\sin(\theta_{m<}-\theta_{min })+\sin(\theta_{m>}-\theta_{min })]=0. \label{21}
\end{align}
The adaptation of (21) is essential to assimilate $\tilde{\Delta}_{max}=(\omega_{m>}-\omega_{min})$ and $\tilde{\Delta}_{m}=(\omega_{m<}-\omega_{max})$ as differences between nearest neighbour oscillators. These guide us, for the second selection (when we take into account the phase terms in (21)), to rewrite (15) as
\begin{align}
&\sin(\theta_{m>}-\theta_{min })=1, \nonumber \\
&\sin(\theta_{max}-\theta_{m>})=1-\frac{3\omega_{m>}}{H^{II-b}(\omega_{m<}-\omega_{max})+(\omega_{m>}
-\omega_{min})}, \nonumber \\
&\sin(\theta_{min}-\theta_{m<})=1+\frac{3 \omega_{min}}{H^{II-b}(\omega_{m<}-\omega_{max})+(\omega_{m>}
-\omega_{min})}, \nonumber \\
%and \nonumber \\
&\sin(\theta_{m<}-\theta_{max})=\frac{H^{II-b}(\omega_{m<}-\omega_{max})+(\omega_{m>}
-\omega_{min})-3(\omega_{max}+\omega_{m>})}{H^{II-b}(\omega_{m<}-\omega_{max })+(\omega_{m>}-\omega_{min})} \nonumber \\
& \equiv \frac{H^{II}-b(\omega_{m<}-\omega_{max})+(\omega_{m>}-\omega_{min})
+3(\omega_{m<}+\omega_{min})}{H^{II-b}(\omega_{m<}-\omega_{max})+(\omega_{m>}-\omega_{min})}, \label{22}
\end{align}
where $H^{II-b}$ is specified in equation (14).

Equations (17), (18), (20) and (22) indicate that the phase differences are defined uniquely within the configurations in each class. In class II, more details are introduced which guide to perceive two cases contingent on the conditions leading to equations (18) and (19). The phase differences, within each class, expose definitely the dependencies on the initial frequencies of the oscillators \{$\omega_{max}>\omega_{m>}>\omega_{m<}>\omega_{min}$\}.

\section{Conclusion}
We have studied a system of four non-identical nearest neighbour bidirectional coupled phase oscillators (BLCKM) in a ring, at the moment the state of a complete frequency synchronization comes out. For this system, we shed the light to the minutiae that arise because the nearest neighbour interactions. These intricacies, in the dissimilar four BLCKM oscillators (a system includes twenty four different arrangements to distribute the starting values of frequencies over the individual oscillators \{$\omega_{max}>\omega_{m>}>\omega_{m<}>\omega_{min}$\} prior to coupling), allow us to categorize three classes. Each class comprises eight configurations. At the synchronization stage, when considering the local coupling between oscillators, we specify a phase lock condition for classes I, II and III. We use the phase lock condition to obtain mathematical expressions, for calculating the critical coupling strengths for any configuration within each class, in the case of four locally coupled phase oscillators in a ring. The expressions for computing the critical coupling strengths, for the three classes, show dependency on the initial frequencies of the four oscillators. Also, we determine the lower and upper limits of the critical coupling constant for each class. We are also capable to obtain formulas for the phase differences, for each class, which demonstrate reliance on the preliminary frequencies of the four nearest neighbour oscillators earlier to coupling.

The analytic expressions, for the critical coupling of non-identical four BLCKM oscillators, will allow to extend the work presented here to study the different synchronization patterns in each class either during partial synchronization or during the unison state for $K > K_c$. Thus, the determination of the coupling factor for non-identical four BLCKM at a phase locking state will assist to obtain formulas for the coupling constants at partial synchronization states. Also, the formulas for $K_c$ will help to calculate the phase differences for a coupling constant larger than the critical coupling value as well as to relate the phase differences to each other. In addition, a further investigation of the four non-local BLCKM can be achieved, for each class, concerning in details the synchronization trees for different distributions of the initial frequencies over the set \{$\omega_{max}>\omega_{m>}>\omega_{m<}>\omega_{min}$\}. A future work can be carried out to examine the transition from a pattern to another within each class. Also, it can introduce a further understanding on the dependency of each pattern, within each class, on the ratio $\tilde{\Delta}_m/\tilde{\Delta}_{max}$. Not only this previous part but also the future work have to explore how the quantity $H$ depends on $\tilde{\Delta}_m/\tilde{\Delta}_{max}$ within each class. The upcoming study must show how the nearest neighbour interaction terms lead to attractions and/or repulsions between oscillators in any arrangements for each class. The previous statement means that we can find a reason for having different critical couplings for the three classes even if the oscillators (within classes I, II and III) have similar initial frequencies. Moreover, a study of a stability analysis can be carried out. A comparative explorations between the four local BLCKM and the four all-to-all coupled model be required because the two cases contain the smallest number of oscillators that introduce the difference between local and global couplings.

It is expected that the study of a case of four non-identical BLCKM oscillators, with periodic boundary conditions, may help to understand the mechanism of synchronization of a few and a limited number of local coupled phase oscillators (unidirectional and bidirectional couplings) in a ring topology. The investigation of the case $N = 4$ BLCKM oscillators could possibly help in interpreting the behavior of systems, for example the case of $N =5$. The similar procedures of the classification of classes can be applied straightforward to the case of $N = 5$ but with more details. Understanding the cases of $N = 4$ and $N = 5$ of nearest neighbour coupled phase oscillators may help to develop a method to study extra number of nearest neighbour coupled oscillators. This development may be done based on what is presented here in this work or based on synthesizing a method that mixes some of ideas in this work and ideas from other works [27, 28]. Also, the presented analysis, in the case of four unalike BLCKM oscillators, may guide us to construct a theory concerning the synchronization of a few and a finite number of non-identical local coupled phase oscillators in a ring. Thus, enforce using a few and a finite number of oscillators to design practical systems that serve in several convenient applications \cite{33}.

Considering the non-identical BLCKM oscillators, the periodic boundary conditions play an extremely important role at the transition to synchronization \cite{28}. However, for a few and a limited number of nearest neighbours coupled phase oscillators, we have to analyze in details the individual oscillators that compose such systems. This extensive study has to be done to determine the two oscillators, which have the phase correlation and draw all of the nearest neighbour oscillators to have the same frequency at a critical coupling.

%%%%%%%%%%%%%%%%%%%%%%%%%%%%%%%%%%%%%%%%%%%%%%%%%%%%%%

%
\end{document}